\documentclass[sn-mathphys-num]{sn-jnl}% Math and Physical Sciences Numbered Reference Style 
\usepackage{graphicx}%
\usepackage{multirow}%
\usepackage{amsmath,amssymb,amsfonts}%
\usepackage{amsthm}%
\usepackage{mathrsfs}%
\usepackage[title]{appendix}%
\usepackage{xcolor}%
\usepackage{textcomp}%
\usepackage{manyfoot}%
\usepackage{booktabs}%
\usepackage{algorithm}%
\usepackage{algorithmicx}%
\usepackage{algpseudocode}%
\usepackage{listings}%

\makeatletter

\providecommand{\tabularnewline}{\\}

\newcommand{\araa}{Annu. Rev. Astron. Astrophys.} % Annual Review of Astron and Astrophys
\newcommand{\apj}{Astrophys. J.} % Astrophysical Journal
\newcommand{\apjl}{Astrophys. J. Lett.} % Astrophysical Journal, Letters
\newcommand{\apjs}{Astrophys. J. Suppl. Ser.} % Astrophysical Journal, Supplement
\newcommand{\aap}{Astron. Astrophys.} % Astronomy and Astrophysics
\newcommand{\cqg}{Class. Quantum Gravity} % Classical and Quantum Gravity
\newcommand{\ijmpd}{Int. J. Mod. Phys. D} % International Journal of Modern Physics D: Gravitation, Astrophysics and Cosmology
\newcommand{\jcap}{J. Cosmol. Astropart. Phys.} % Journal of Cosmology and Astroparticle Physics
\newcommand{\jatis}{J. Astron. Telesc. Instrum. Syst.} % Journal of Astronomical Telescopes, Instruments, and Systems
\newcommand{\mnras}{Mon. Not. R. Astron. Soc.} % Monthly Notices of the RAS
\newcommand{\mnrasl}{Mon. Not. R. Astron. Soc. Lett.} % Monthly Notices of the RAS
\newcommand{\prd}{Phys. Rev. D} % Physical Review D
\newcommand{\prl}{Phys. Rev. Lett.} % Physical Review Letters
\begin{document}

\title[Dipole in line-galaxy number counts and line intensity maps]{Kinematically Induced Dipole Anisotropy in Line-Emitting Galaxy Number Counts and Line Intensity Maps}

\author*{\fnm{Kyungjin} \sur{Ahn}}\email{kjahn@chosun.ac.kr}

\affil{\orgdiv{Department of Earth Sciences}, \orgname{Chosun University}, \orgaddress{\city{Gwangju}, \postcode{61452}, \country{Korea}}}

\abstract{The motion of the solar system against an isotropic radiation background,
such as the cosmic microwave background,
induces a dipole anisotropy in the background due to the Doppler effect.
Flux-limited observation of the continuum radiation from galaxies
also has been studied extensively to show a dipole anisotropy due
to the Doppler effect and the aberration effect. We show that a similar
dipole anisotropy exists in spectral-line intensity maps, represented
as either galaxy number counts or the diffuse intensity maps. The
amplitude of these dipole anisotropies is determined by not only the
solar velocity against the large-scale structures but also the temporal
evolution of the monopole (sky-average) component. Measuring the dipole
at multiple frequencies, which have mutually independent origins due
to their occurrence from multiple redshifts, can provide a very accurate
measure of the solar velocity thanks to the redundant information.
We find that such a measurement can even constrain astrophysical parameters in the nearby universe.
We explore the potential for dipole measurement of existing and upcoming
surveys, and conclude that the spectral number count of galaxies through SPHEREx
will be optimal for the first measurement of the dipole anisotropy in the spectral-line galaxy
distribution. LIM surveys with reasonable accuracy are also found to be promising.
We also discuss
whether these experiments might reveal a peculiar nature of our local
universe, that seems to call for a non-standard cosmology other than
the simple $\Lambda$CDM model as suggested by recent measures of the
baryon acoustic oscillation signatures and the Alcock-Paczynski tests.}

\keywords{Cosmology, Cosmic microwave background, Dark energy, Sky Surveys}

\maketitle

\section{Introduction}\label{sec:Introduction}

Among the anisotropic modes in one of the most studied continuum foregrounds,
the cosmic microwave background (CMB), the dipole anisotropy stands
out in terms of its different origin from the rest (quadrupole and
higher-order multipoles). The dipole anisotropy of the CMB in temperature
is about $1.23\times10^{-3}$ of the monopole value (2.725K), and thus
is the largest of all the multipoles of the CMB anisotropy that, except
for the dipole moment, range at $\lesssim10^{-5}$ of the monopole.
Such a largeness of the dipole anisotropy relative to other multipoles
indicates that the dipole has a different origin, which is believed to be caused by the relative
velocity of the observer (solar system) against the rest frame of
CMB. It is thus natural to ignore the intrinsic dipole anisotropy
due to the fluctuation in structure formation but instead fully attribute
the CMB dipole to the kinematic origin and estimate the solar velocity $v_{\odot}$ in units of $c$, 
$\beta_{\odot}\equiv v_{\odot}/c=(1.23\pm0.017)\times10^{-3}$, against
the CMB-rest frame. Of course the intrinsic dipole anisotropy must exist, and may be separated out if its small impact on the spectra of the monopole and quadrupole moments can be detected \cite{Yasini2017}. 

Dipole anisotropy exists also in the number distribution of continuum-emitting
radiation sources, if observed with a fixed flux limit. Such a dipole
must arise from the same kinematic origin, because the Doppler effect
boosts the flux of galaxies toward the direction of the solar motion and
thus increases the number of galaxies detected above a given flux limit.
In addition, the light aberration also contributes to the dipole because
galaxies will look more clustered toward the direction of motion than the opposite direction.
This phenomenon was first formulated by \cite{Ellis1984}, who worked
on the case when continuum radiation from galaxies had a power-law
flux $F\propto\nu^{-\alpha}$ and so does the number density, $N(>F)\propto F^{-x}$,
to find that the amplitude of the dipole ($D$) becomes $D=[2+x(1+\alpha)]\beta_{\odot}$.
Subsequently, actual observations of the dipole anisotropies in the number density of
continuum-emitting galaxies \cite{Xia2010,Singal2011,Chen2016}
and quasars \cite{Secrest2021} followed to find that the estimated
$\beta_{\odot}$ is $\sim$2--5 times as high as that deduced from
the CMB dipole, which is a very puzzling conflict for a single quantity. This
conflict is considered by some as one of the non-negligible ``tensions''
between the standard cosmological model and observations, which mostly
reside in the local universe of the line-of-sight comoving distance
$\lesssim3\,{\rm Gpc}$ or $z\lesssim1$ \cite{Dalang2022,KumarAluri2023}.

There are about three categories of resolutions to this apparent paradox.
First, systematic uncertainties arising from observations themselves
are to blame. Re-analysis of the continuum galaxy data of the WISE
(Wide-field Infrared Survey Explorer) survey resulted in $\beta_{\odot}$
estimation statistically consistent with the CMB-based one \cite{Darling2022},
in contrast to the same WISE-based analysis undertaken previously \cite{Secrest2021},
attributing the conflict to the systematic uncertainty caused by the
observing and data-collecting schemes in WISE. However, the trend
of estimated $\beta_{\odot}$ from galaxy counts being different from
the CMB-based estimation seems still persistent \cite{Panwar2024}.
Even the direction of the dipole is in conflict, and thus
some physical origin other than the kinematic origin may be responsible. Second,
there may be an additional contribution to the dipole anisotropy from
the biased structure formation. As suggested by \cite{Tiwari2016},
a high level of galaxy bias of $\sim3$ could significantly boost
the galaxy formation in high-density region, which is also a gravitational
attractor driving the solar motion. However, the directional discrepancy
between the galaxy number dipole and the CMB dipole (e.g. \cite{Panwar2024})
seems to invalidate such claim. Third, one could explain such a discrepancy
by adopting a non-standard cosmology with an anomalous universe. Then,
one could have the CMB-rest frame in a relative motion against the
matter-rest frame, resulting in mismatching dipole anisotropies probed
by the CMB and the matter-density probes, namely galaxies \cite{Domenech2022}. 

Faced with the puzzle, more independent measures of similar dipole
anisotropies seem necessary. Here we suggest using spectroscopic galaxy
surveys (SGS) and line intensity mappings (LIM) to obtain their
dipole anisotropy. To our knowledge, no such studies have been conducted
to date, neither in theory or in observation.
The SGS is to construct 3-dimensional galaxy maps from
discrete, individually identifiable sources with redshift indicators such as a spectral line.
The  LIM is to construct 3-dimensional intensity maps from sources that
are not identifiable --- protogalaxies, galaxies and the intergalactic
medium --- by a telescope but collectively forming a diffuse background.
However, so far all such surveys are either limited in the
sky coverage, or aiming at the full sky but without spectroscopic capabilities.
The only full-sky spectroscopic survey is SPHEREx (Spectro-Photometer
for the History of the Universe, Epoch of Reionization, and Ices Explorer,
\cite{Crill2020}), which will be launched in the near future.
In the dipole measurement, SGS and LIM can
be more beneficial than the continuum-galaxy mapping because one could obtain
true redundancy in estimating $\beta_{\odot}$ by probing multiple redshifts
through multiple observing frequencies. In contrast, even though the
continuum-galaxy mapping has multi-frequency information, the amplitude of
the continuum spectrum at any given frequency is an integral of multi-redshift
contributions and thus lacks the distinctiveness of origin (and the
corresponding redundancy) seen in line-galaxy surveys and LIMs.
For example, huge radio interferometers have been in operation or
are being built to map the intensity of the hydrogen 21-cm line (e.g.
LOFAR (LOw Frequency ARray): \cite{Mertens2020}, SKA-LOW
(Square Kilometre Array -- Low frequency): \cite{Ahn2015b,Labate2022}),
either from very high redshifts ($z\gtrsim6$ or the observing frequency
$\nu\lesssim200\,{\rm MHz}$) or from relatively low redshifts ($0\le z\lesssim6$
or $200\,{\rm MHz}\lesssim\nu\le1.4\,{\rm GHz}$). We have elsewhere
forecasted observations of the large-angle anisotropy
of 21-cm lines: the dipole anisotropy \cite{Hotinli2024}
and the integrated Sachs-Wolfe effect \cite{Ahn2024} in the 21-cm background,
both being the largest-angle (solid angle $\gtrsim(20^{\circ})^{2}$)
phenomena and thus requiring almost full-sky LIMs. We intend to extend and generalize
our study of the 21-cm dipole anisotropy to the SGS and LIM dipoles.

This paper is organized as follows. In Sect. \ref{sec:Theory},
we calculate the amplitude of the dipole anisotropy in SGS and LIM.
In Sect. \ref{sec:Forecasts}, we perform forecasting possible observations
of SGS and LIM dipoles. Sect. \ref{sec:Discussion} is dedicated
to a brief summary and the discussion on the prospects of these proposed
observations, with an emphasis on a serendipitous case when the estimated
solar velocity happens to be found inconsistent with the value estimated by the
CMB dipole measurement.

\section{Theory}\label{sec:Theory}

\subsection{Notations and basic relations}

We start from the cosmological principle asserting that the universe
is homogeneous and isotropic in largest scales. The dipole anisotropy
is of course the largest-scale anisotropy, and thus is the case this
principle suits best. We then need to introduce two different reference
frames: (1) the background-rest frame (BRF) and (2) the observer-rest
frame (ORF). The BRF denotes a frame in which the observer is comoving
with the expansion of the universe, and to this observer the universe
seems isotropic including any radiation background. The ORF denotes
a frame attached to the solar system having us as the observer, and
as described in Sect. \ref{sec:Introduction} a radiation background
will show the kinematically induced anisotropy. 

We adopt a convention in which any scalar quantity $A$ and vector
quantity ${\bf a}$ will be denoted without a prime in BRF while with
a prime in ORF ($A'$ and ${\bf a}'$). For example, $F_{\nu}$ represents
the (true) differential flux at the observing frequency $\nu$ for
an observer in BRF, while $F'_{\nu'}$ represents the (apparent) differential
flux at the observing frequency $\nu'$ for an observer in ORF.
A quantity in one frame with an argument in another can also be defined:
e.g. $F_{\nu'}$ represents the (true) differential flux but measured
at $\nu'$ by an observer in ORF. Such a convention is a common practice
especially in the literature of the kinematically induced anisotropy
(e.g. \cite{Bottani1992,Hotinli2024}). And we use the ``apparent''
cosmological redshift $z'$ in ORF measured through spectroscopy of a
line, such that $1+z'=\nu_{0}/\nu_{{\rm c}}'$ if the measured line
center is at frequency $\nu_{{\rm c}}'$. From this point on, we assume
all measurements are made at fixed $z'$ (and with a fixed set of
bandwidths) regardless of the line-of-sight direction $\hat{n}'$.
Due to the smallness of $\beta_{\odot}$, one can expand the observed field
in multipoles with ORF quantities in decreasing order, and up to the
dipole we denote
\begin{equation}
A'(z',\hat{n}')=A'(z')+A'_{{\rm dip}}\mu'=A(z')+A'_{{\rm dip}}\mu',\label{eq:dipoledefinition}
\end{equation}
where $\mu'\equiv\cos\theta'$ with the azimuthal angle $\theta$ against the direction of
the observer's motion,
and $A'(z')=A(z')$ is the monopole component due to sources at redshift
$z'$\footnote{There exists $\mathcal{O}(\beta_{\odot}^{\ge 2})$ correction on the monopole
component in ORF \cite{Challinor2002,Hotinli2024}, due to the bleeding of the monopole into other multipoles.
However, this correction is small and thus validates Eq. (\ref{eq:dipoledefinition}) up to $\mathcal{O}(\beta_{\odot})$.}. 

Lorentz transformation gives the following conversion rules between
quantities in BRF and those in ORF \cite{Rybicki_Lightman,Bottani1992,Hotinli2024}:
\begin{align}
\mu' & =\frac{\mu+\beta}{1+\beta\mu}\nonumber \\
\frac{\nu'}{\nu}= & \frac{d\nu'}{d\nu}=\frac{1}{\gamma(1-\beta\mu')},\nonumber \\
\frac{dt'}{dt}= & \gamma(1-\beta\mu'),\nonumber \\
\frac{d\Omega'}{d\Omega}= & \gamma^{2}(1-\beta\mu')^{2},\nonumber \\
\frac{F'}{F}= & \frac{dF'}{dF}=\frac{1}{\gamma^{2}(1-\beta\mu')^{2}},\label{eq:conversion}
\end{align}
where $\beta$ is the velocity (in units of $c$) of the observer
against BRF, $dt'$ is the photon arrival interval (thus the extra
$1-\beta\mu'$ in addition to the time dilation factor $\gamma=(1-\beta^{2})^{-1/2}$:
\cite{Rybicki_Lightman} pp. 141), $F$ is the bolometric flux of a line
from a point source (e.g. galaxy) such that $F_{\nu}\equiv dF/d\nu =F\phi(\nu)$
with the intrinsic line profile $\phi$ ($\int\phi(\nu)\,d\nu=1$, and we
assume $\phi(\nu)\simeq\delta(\nu-\nu_{0})$ with the Dirac delta
function $\delta$ without losing much generality throughout this
paper), $d\Omega$ is the infinitesimal solid angle along the line-of-sight
direction $\hat{n}$. For a source residing at cosmological redshift
$z$, the redshifted line center is at $\nu_{{\rm c}}=\nu_{0}/(1+z)$,
while an ORF observer will have a shifted redshift $z'$ satisfying
\begin{align}
\nu_{{\rm c}}'= & \nu_{0}/(1+z'),\nonumber \\
1+z'= & \gamma(1-\beta\mu')(1+z)\label{eq:zprime}
\end{align}
even though it is not too common a practice to correct the observed
redshift $z'$ with Eq. (\ref{eq:zprime}).

\subsection{Dipole anisotropy in galaxy number count}\label{subsec:galaxynumber}

We construct a monopole ``field'' out of the point sources which
are presumed to be isotropic in distribution and with universal physical
quantities at a given redshift, for a comoving observer. The sources are assumed to have redshift
indicators, or at least one spectral line. For a specific line emission (e.g. HI Ly$\alpha$),
we can again define the bolometric line flux $F$ as $F_{\nu} =F\phi(\nu)$. Hereafter, $F$ denotes
this bolometric line flux.

The number density of line-emitting sources will bear a dipole anisotropy as
follows. Suppose, in BRF, there are total $N$ galaxies in the frequency
range $\nu=[\nu_{{\rm min}},\,\nu_{{\rm max}}]$ and the flux range
$F=[F_{{\rm min}},\,F_{{\rm max}}]$. Note that the observing frequency $\nu$ corresponds
to the cosmological redshift $z$, and thus galaxies observed within some frequency range are
those residing in the corresponding redshift range.
Then, we can define the (differential)
number density (galaxy number per unit frequency, flux
and solid angle) as
\begin{equation}
n(F,\,\nu)=\frac{dN}{d\nu\,dF\,d\Omega},\label{eq:numdensity}
\end{equation}
where we assumed isotropy in BRF and thus $n$ does not have dependence
on $\hat{n}$\footnote{If wanted, one can always use a new quantity $n(F,\,z)\equiv dN/dz\,dF\,d\Omega=n(F,\,\nu)(1+z)/\nu.$}.
From now on, we assume that the universe is isotropic in BRF
for the sake of calculating the kinematic dipole anisotropy.
Even after the Lorentz transformation, the Lagrangian
number should be conserved, or $dN'=dN$, satisfying
\begin{equation}
n'(F',\,\nu',\,\hat{n}')=n(F,\,\nu)\frac{d\nu}{d\nu'}\frac{dF}{dF'}\frac{d\Omega}{d\Omega'}=n(F,\,\nu)\gamma(1-\beta\mu'),\label{eq:ndenstrans}
\end{equation}
where the final relation is due to Eq. (\ref{eq:conversion}).
After Taylor-expanding $n(F,\,\nu)$ in terms of the ORF parameters
$F'$ and $\nu'$ in Eq. (\ref{eq:ndenstrans}), and using Eq.
(\ref{eq:zprime}), we obtain the dipole moment of $n'$:
\begin{equation}
n'_{{\rm dip}}(F',\,\nu')=-n(F',\,\nu')\left(1+2\left.\frac{\partial\ln n}{\partial\ln F}\right|_{F',\nu'}+\left.\frac{\partial\ln n}{\partial\ln\nu}\right|_{F',\nu'}\right)\beta.\label{eq:n}
\end{equation}
In practice, one might want to use the cumulative number density 
\begin{equation}
\mathcal{N}(\nu)\equiv\frac{dN}{d\nu\,d\Omega}=\int dF\,n(F,\,\nu).\label{eq:Numberdensity}
\end{equation}
This would be useful when the number of objects probed by a survey
is not large enough to reliably construct $n'$, and thus it becomes necessary to mitigate the shot noise.
Using the fact that $dN'=dN$ or $\mathcal{N}'(\nu',\hat{n})\,d\Omega'd\nu'=\mathcal{N}(\nu)\,d\Omega d\nu$
with the help of Eq. (\ref{eq:conversion}), we find that
\begin{equation}
\mathcal{N}'_{{\rm dip}}(\nu')=\mathcal{N}(\nu')\left(1-\left.\frac{\partial\ln\mathcal{N}}{\partial\ln\nu}\right|_{\nu'}\right)\beta.\label{eq:N}
\end{equation}

There are pros and cons in using $n'$ and $\mathcal{N}'$ for the
dipole measurement. If we use $n'$, the dipole depends on $F'$ which
is of the astrophysical origin such as the mass-to-light ratio of
cosmological halos, star formation rate, black-hole formation rate,
etc. Therefore, the dipole of $n'$ will bear astrophysical information
just as the monopole of $n'$, in addition to the information on $\beta_{\odot}$.
However, this will be realized only when the number of galaxies probed
is large enough to span some flux range, because otherwise the statistical
uncertainty from the shot noise might dominate the true signal. If
we instead use $\mathcal{N}'$, the required number of objects
to probe would not be as severe as for $n'$. However, the astrophysical
information will be only limited to the cumulative number in frequency
bins, or equivalently redshift bins, and thus the information content
will be more limited. Nevertheless, as long as our interest lies only
in probing $\beta_{\odot}$, the dipole of $\mathcal{N}'$ could provide
multiple frequency observables to constrain $\beta_{\odot}$. We show
in Sect. \ref{subsec:galaxy-number-count} our forecast
that can be used as a guide to future galaxy surveys.

Note that the derivation of Eqs. (\ref{eq:n}) and (\ref{eq:N}) are almost identical to that by Ref. \cite{Dalang2022} in that the calculation is based off the Lagrangian conservation of the number count across the BRF and the ORF. The difference lies in the fact that our derivation is for line emission but that by Ref. \cite{Dalang2022} is for continuum emission.

\subsection{Dipole anisotropy in line intensity map}\label{subsec:LIM}

Line intensity maps at a certain observing frequency $\nu$ is associated
with a collection of lines emitted (or absorbed, such as the Ly$\alpha$
forest) by various astrophysical objects at the cosmological redshift
$z$ satisfying $\nu=\nu_{0}/(1+z)$ with the line central frequency
$\nu_{0}$. In BRF, the specific intensity (net energy per observed
frequency, time, area and solid angle) is given by the cosmological
radiative transfer equation,

\begin{equation}
I_{\nu}=\int ds''\frac{j_{\nu''}(z'')}{(1+z'')^{3}},\label{eq:intensity-basic}
\end{equation}
where $ds$ is the light-of-sight distance travelled by a photon,
$j_{\nu}\equiv dE/dV\,d\Omega\,dt\,d\nu$ with the energy $E$ and
the \emph{proper} volume $V$, is the \emph{proper} monochromatic
emission coefficient (\cite{Rybicki_Lightman}, pp. 9) in the source-rest
frame, and the factor $1/(1+z)^{3}$ arises from the effect of the
cosmological redshift on the observable (e.g. \cite{Peacock1999},
Eq. 3.88). The absorption of a line by any gas in between the source
and the observer, either in BRF or ORF, does not exist in our setup
because every gas particle is assumed to be comoving and thus is always
away from the redshifted line center. $j_{\nu}$ of line sources in
terms of the \emph{comoving} luminosity density $\rho_{{\rm c}}\equiv dL/dV_{{\rm c}}=(1+z)^{3}dL/dV$,
with the comoving volume $V_{{\rm c}}$ and the emitted frequency
$\nu_{{\rm e}}$, from sources at $z$ is given by
\begin{equation}
j_{\nu_{{\rm e}}}(z)=\frac{(1+z)^{3}\rho_{{\rm c}}(z)}{4\pi}\phi(\nu_{{\rm e}})=\frac{(1+z)^{3}\rho_{{\rm c}}(z)}{4\pi}\delta(\nu_{{\rm e}}-\nu_{0}).\label{eq:emission}
\end{equation}
Using Eq.s (\ref{eq:intensity-basic}) and (\ref{eq:emission})
together with relations $\nu=\nu_{{\rm e}}/(1+z)$ and $ds=dz\,c/[H(z)(1+z)]$,
we obtain
\begin{equation}
I_{\nu}=\frac{c\rho_{{\rm c}}(z)}{4\pi\nu_{0}H(z)},\label{eq:intensity}
\end{equation}
where $\nu$ is now implicitly related to $z$ as $\nu=\nu_{0}/(1+z)$. 

Now we can address the dipole anisotropy of LIM using the conservation
law $I'_{\nu'}/I_{\nu}=(\nu'/\nu)^{3}$ under the Lorentz transformation,
or the Liouville theorem on the specific intensity (e.g. \cite{Rybicki_Lightman}).
Using the Liouville theorem and Taylor-expanding $I_{\nu}$ in terms
of quantities in ORF with the help of Eq. (\ref{eq:conversion}),
we obtain
\begin{equation}
\frac{I'_{\nu',{\rm dip}}}{I_{\nu'}}=\left(3-\left.\frac{\partial\ln I_{\nu}}{\partial\ln\nu}\right|_{\nu'}\right)\beta=\left(3+\left.\frac{\partial\ln\rho_{{\rm c}}}{\partial\ln(1+z)}\right|_{z'}-\left.\frac{\partial\ln H}{\partial\ln(1+z)}\right|_{z'}\right)\beta.\label{eq:Inudipole}
\end{equation}
Eq. (\ref{eq:Inudipole}) shows two aspects: (1) the first equality
shows that the dipole of $I'_{\nu'}$ is determined by $\beta$ and
the the monopole $I_{\nu'}$ (amplitude and the spectral shape), and
(2) the second equality shows a further dependence on cosmology through
its dependence on the Hubble parameter. Even though the latter aspect
seems attractive, using the dipole for probing $H$ and its underlying
parameters such as the matter content ($\Omega_{m}$) and the dark
energy equation of state ($\rho_{{\rm DE}}\propto a^{-1(3+w)}$) can
be obtained only when prior knowledge on $\rho_{{\rm c}}(z')$ is
established with the help of e.g. standard candles. Otherwise, we
can only measure $I_{\nu'}$, and degeneracy between $\rho_{{\rm c}}(z')$
and $H(z')$ is unavoidable for whichever cosmology we assume.
$\rho_{{\rm c}}(z')$ is to $I_{\nu'}$ what luminosity is to flux, and thus
without a prior knowledge on the luminosity distance ($D_L$),
$\rho_{{\rm c}}(z')$ cannot be deduced from a sheer observation of $I'_{\nu'})_{\rm dip}/I_{\nu'}$.
Diffuse LIMs do not have the luxury of hosting well-defined standard
candles and are dominated by collection of lights from otherwise undetectable
sources, and therefore the potential of using the LIM dipole for cosmology
is slim.

Instead, the dipole of LIM can be used to perform astrophysics on
large scales as follows. First, because there exist several independent
estimates of $H(z')$, we can use these as a prior on $H(z')$ and
then estimate $\beta_{\odot}$ and the temporal evolution of $\rho_{{\rm c}}$,
or $\partial\ln\rho_{{\rm c}}/\partial\ln(1+z)|_{z'}$, through observing
$I'_{\nu',{\rm dip}}/I_{\nu'}$. This information can
then be linked to astrophysics such as the evolution of the star formation
rate (e.g. Ly$\alpha$ LIM) or the amount of neutral gas (e.g. 21-cm
LIM), etc. Second, under a given prior both on $H(z')$ and $\beta_{\odot}$,
we could quantify the intrinsic dipole anisotropy or find a tension
in the measure of $\beta_{\odot}$. Suppose that we first measure
$\rho_{{\rm c}}(z')$ through the measure of the monopole $I_{\nu'}$
and the prior on $H(z')$, using Eq. (\ref{eq:intensity}). Then,
the measure of the dipole $I'_{\nu',{\rm dip}}/I_{\nu'}$
should satisfy the last equality of Eq. (\ref{eq:Inudipole}),
because all variables are constrained ($\rho_{{\rm c}}$ from the
monopole measure, and $H$ and $\beta_{\odot}$ from the prior). However,
if the measure of the dipole $I'_{\nu',{\rm dip}}/I_{\nu'}$
does not satisfy this equality, this means either the prior on $\beta_{\odot}$ (e.g. from CMB dipole)
is wrong or there exists an intrinsic dipole moment due to the inhomogeneous
distribution of radiation sources projected as the anisotropy in the
LIM. Relative motion of the matter-rest frame against the radiation-background-rest frame
could be another resolution for such $\beta_{\odot}$ tension.

\begin{figure}[h]
\centering
\includegraphics[width=0.6\textwidth]{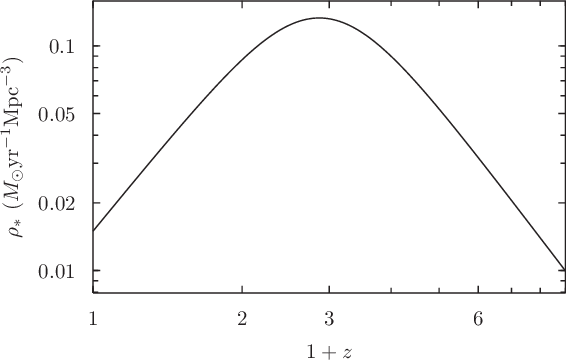}
\caption{Evolution of the star formation rate density shown as a fit (Eq.~\ref{eq:rhostar})
to the observed data \cite{Madau2014}.}\label{fig:rhostar}
\end{figure}

We adopt a parametric approach for inference. First, we assume a 3-parameter
fit to the evolution of $\rho_{{\rm c}}$ such that
\begin{equation}
\frac{\partial\ln\rho_{{\rm c}}}{\partial\ln(1+z)}=\gamma_{1}-\frac{(\gamma_{1}-\gamma_{2})(1+z)^{\gamma_{1}-\gamma_{2}}}{(1+z_{{\rm t}})^{\gamma_{1}-\gamma_{2}}+(1+z)^{\gamma_{1}-\gamma_{2}}}.\label{eq:dlrhodz1}
\end{equation}
Eq. (\ref{eq:dlrhodz1}) is simply to test the power of the dipole
measurement: the specific form used in Eq. (\ref{eq:dlrhodz1})
can sometimes pose a strong prior effect, in which case one can change
the functional form into a more appropriate one if some hint on $\rho_{{\rm c}}$
is available. Nevertheless, Eq. (\ref{eq:dlrhodz1}) is based
on the evolution of the star formation rate density (stellar mass
generated per time and comoving volume) $\rho_{*}$ at $z\lesssim8$,
well fit by the form
\begin{equation}
\rho_{*}(z)=\frac{2\rho_{*}(z_{{\rm t}})}{(1+z_{{\rm t}})^{\gamma_{1}}}\frac{(1+z)^{\gamma_{1}}}{1+[(1+z)/(1+z_{{\rm t}})]^{\gamma_{1}-\gamma_{2}}}.\label{eq:rhostar}
\end{equation}
$\rho_{*}$ in this form peaks at $z\simeq z_{{\rm t}}$, and roughly
follows a broken power law: $\rho_{*}(z<z_{{\rm t}})\propto(1+z)^{\gamma_{1}}$
and $\rho_{*}(z>z_{{\rm t}})\propto(1+z)^{\gamma_{2}}$. The observed
$\rho_{*}$ is well fit by \{$\rho_{*}(z_{{\rm t}})$, $z_{{\rm t}}$,
$\alpha$, $\beta$\}$\simeq$\{0.133$\,M_{\odot}{\rm yr}^{-1}{\rm Mpc}^{-3}$,
1.9, 2.7, -2.9\} in a wide redshift range, $z$=0--8 \cite{Madau2014}. See Fig.~\ref{fig:rhostar} for this fit.
If one assumes a constant mass-to-light ratio M/L, it is reasonable
to let $\rho_{{\rm c}}\propto\rho_{*}$ and thus we adopt Eq.
(\ref{eq:dlrhodz1}) in this work. In order to accommodate the dependence
of $H$ on cosmological parameters when using a prior on cosmology,
we use the fact that $\partial\ln H/\partial\ln(1+z)=\partial\ln E/\partial\ln(1+z)$
with
\begin{equation}
E(z)\equiv H(z)/H_{0}=\left[\Omega_{m,0}(1+z)+\Omega_{k,0}(1+z)^{2}+(1-\Omega_{m,0}-\Omega_{k,0})f(z,w)\right],\label{eq:Ez}
\end{equation}
where $w$ is the dark energy equation of state,
\begin{equation}
f(z)=\exp\left(3\int_{0}^{z}\frac{1+w(z'')}{1+z''}dz''\right)=(1+z)^{3(1+w_{0}+w_{a})}\exp\left(-3w_{a}\frac{z}{1+z}\right),\label{eq:fz}
\end{equation}
and the last equality of Eq. (\ref{eq:fz}) is valid only when
one adopts the CPL (Chevallier-Polarski-Linder) parametrization of $w(z)$ \cite{Chevallier2001,Linder2003}, namely
\begin{equation}
w(z)=w_{0}+w_{a}\frac{z}{1+z}.\label{eq:CPL}
\end{equation}

\section{Forecasts}\label{sec:Forecasts}

\subsection{Galaxy number count}\label{subsec:galaxy-number-count}

The dipole anisotropy in the galaxy number count is affected by the
number of probed galaxies and the uncertainty in redshifts of a spectral
line. The former and the latter restrictions mainly come from the
telescope sensitivity and the spectral resolution, respectively. As
the Poisson shot noise from a finite number of galaxies is one of
the most severe restrictions, we focus only on $\mathcal{N}'(\nu',\hat{n})$
that could warrant the shot noise smaller than $n'(F',\,\nu',\,\hat{n}')$.
The simple form of Eq. (\ref{eq:N}) allows an easy analytical
assessment of the uncertainty at a frequency bin, by propagating errors,
\begin{equation}
\sigma_{\beta}(\nu')=\frac{\sqrt{\mathcal{N}'^{2}\sigma_{D}^{2}+D^{2}\sigma_{\mathcal{N}'}^{2}}}{\mathcal{N}'^{2}\left|1-\left.\frac{\partial\ln\mathcal{N}}{\partial\ln\nu}\right|_{\nu'}\right|},\label{eq:sigmabeta}
\end{equation}
where $\mathcal{N}'\equiv\mathcal{N}'(\nu')$, $D\equiv\mathcal{N}'_{{\rm dip}}$
and $\sigma_{A}$ is the standard deviation of a quantity $A$. Each
uncertainty is inclusive of possible systematic uncertainties: e.g.
$\sigma_{\mathcal{N}'}=(\mathcal{N}'+\sigma_{{\rm ext}}^{2})^{1/2}$
where $\mathcal{N}'^{1/2}$ is the Poisson shot noise and $\sigma_{{\rm ext}}$
is the uncertainty caused by the limited spectral resolution, etc.

Assuming an ideal case where the shot noise is the only cause of the
uncertainties, one could estimate a most optimistic error (inverse
of the signal-to-noise ratio) by letting $\sigma_{\mathcal{N}'}\simeq\sigma_{D}\simeq\mathcal{N}'^{1/2}$ in Eq. (\ref{eq:sigmabeta}):
\begin{align}
  \frac{\sigma_{\beta}(\nu')}{\beta}&\simeq\frac{1}{\beta\sqrt{\mathcal{N}'\left|1-\left.\frac{\partial\ln\mathcal{N}}{\partial\ln\nu}\right|_{\nu'}\right|}}\nonumber\\
  &\simeq0.36\left(\frac{\mathcal{N}'}{10^{6}}\right)^{-\frac{1}{2}}\left(\frac{\left|1-\left.\frac{\partial\ln\mathcal{N}}{\partial\ln\nu}\right|_{\nu'}\right|}{5}\right)^{-\frac{1}{2}}\left(\frac{\beta}{1.23\times10^{-3}}\right)^{-1}.\label{eq:sigmabetamin}
\end{align}
Eq. (\ref{eq:sigmabetamin}) already puts a severe constraint
on a survey requirement. In order to obtain about $\le36\%$ error
on each redshift bin, one needs to probe at least $\mathcal{N}'\ge10^{6}$
galaxies per the redshift bin with the help of a steep spectral change
in $\mathcal{N}'$, $\left|1-\left.\frac{\partial\ln\mathcal{N}}{\partial\ln\nu}\right|_{\nu'}\right|\simeq5$.
In an unfortunate case with $\left|1-\left.\frac{\partial\ln\mathcal{N}}{\partial\ln\nu}\right|_{\nu'}\right|\le1$,
the error increases to $\ge81\%$ with $\mathcal{N}'=10^{6}$ or could
remain at $\sim36\%$ only when probing more galaxies, $\mathcal{N}'\simeq2.3\times10^{6}$.
When there are $N_{z}$ redshift bins, of course, the net error on $\sigma_{\beta}/\beta$
will decrease to $\sim1/N_{z}^{-1/2}$ of the bin-wise error, Eq.
(\ref{eq:sigmabetamin}).

While demanding, such a requirement is within the easy reach of the
SPHEREx survey that is scheduled to be launched around April 2025
and expected to probe $4.5\times10^{8}$ galaxies residing at $z=0-2$
with spectroscopy. The existing full-sky galaxy surveys with photometry,
such as WISE and 2MASS (Two Micron All Sky Survey) are too marginal
to reliably use the spectral lines because (1) there are only 4 and
3 frequency bands in WISE and 2MASS, respectively, to falter a reliable
spectroscopy and (2) the overall numbers of galaxies probed are only
about $10^{6}$ and $1.6\times10^{6}$ in WISE and 2MASS, respectively.
Note that the dipole measurements using the WISE galaxies utilized
not the line flux but the average continuum flux, which allowed the
usage of the total number of galaxies ($\sim10^{6}$) and allowed
a $\sim50\%$ error on estimating $\beta$ at 1$\sigma$ level. The
spectroscopy-capable SPHEREx, combined with the largeness of the expected galaxy
count, will enable a reliable measure of the dipole anisotropy. Nevertheless, even with SPHEREx the impact of the unavoidable mask on the galactic plane would not be trivial due to the leakage of other multipoles into the dipole and the shrinkage of the sky coverage, which should be carefully addressed in estimating the true dipole \cite{Abghari2024}.

\subsection{Line intensity mapping}\label{subsec:Line-intensity-mapping}

There does not exist a full-sky LIM survey yet other than LIM surveys
on limited regions of the sky, and therefore we perform a forecast
on a possible future full-sky LIM survey. At the moment, intensive
observational work is dedicated to the study of important spectral
lines such as HI Ly$\alpha$ \cite{Oestlin2014,Ahumada2020},
21-cm \cite{Mertens2020,Labate2022,HERA2023}, and CO lines \cite{Keating2016,Li2016},
but rather focusing on a limited target area of the sky. Therefore,
we seriously suggest that low-angular resolution, large-angle LIM
surveys be performed in the near future.

A full-sky LIM survey can be analyzed just by itself or in combination
with other preexisting cosmological observations. Toward this, we
perform a Fisher matrix analysis, using a MATLAB module ''Fisher4Cast''
developed by \cite{Bassett2011}. Fisher4Cast calculates the Fisher
matrix
\begin{align}
F_{AB}=\sum_{\alpha} & \left[\frac{\partial{\bf X}_{\alpha}^{T}}{\partial\theta_{A}}\mathbb{C}_{\alpha}^{-1}\frac{\partial{\bf X}_{\alpha}^{T}}{\partial\theta_{B}}+\frac{1}{2}{\rm Tr}\left(\mathbb{C}_{\alpha}^{-1}\frac{\partial\mathbb{C}_{\alpha}}{\partial\theta_{A}}\mathbb{C}_{\alpha}^{-1}\frac{\partial\mathbb{C}_{\alpha}}{\partial\theta_{B}}\right)\right]\nonumber \\
= & \sum_{\alpha}\sum_{i}\frac{1}{\sigma_{\alpha,i}^{2}}\frac{\partial X_{\alpha}(z_{i})}{\partial\theta_{A}}\frac{\partial X_{\alpha}(z_{i})}{\partial\theta_{B}},\label{eq:fisher}
\end{align}
where an observable ${\bf X}_{\alpha}$ is composed of multi-redshift
probes at \{$z_{i}$\} such that
\begin{equation}
{\bf X}_{\alpha}=\left(X_{\alpha}(z_{1}),\,X_{\alpha}(z_{2}),\,\cdots,\,X_{\alpha}(z_{n})\right),\label{eq:X}
\end{equation}
$\mathbb{C}_{\alpha}$ is the covariance matrix of observable $X_{\alpha}$
such that $(\mathbb{C}_{\alpha})_{ij}\equiv\langle(X_{\alpha}(z_{i})-\langle X_{\alpha}(z_{i})\rangle)(X_{\alpha}(z_{j})-\langle X_{\alpha}(z_{j})\rangle)$,
and $\theta_{i}$ denotes underlying parameters: energy contents ($\Omega_{m,0}$
and $\Omega_{k,0}$), equation of state parameters ($w_{0}$ and $w_{a}$),
the current-day Hubble constant $H_{0}$, the solar velocity against
the background-rest frame $\beta_{\odot}$, and the astrophysical
parameters $\gamma_{1}$, $\gamma_{2}$, and $z_{t}$. The last equality
in Eq. (\ref{eq:fisher}) is usually used, which is in principle
valid only when measurement errors are independent of cosmological
parameters and across observing redshifts. We follow this convention
in this work. When to apply a prior information on $\theta_{i}$,
a diagonal matrix $P_{AB}=\delta_{AB}/\sigma_{A}^{2}$ is added to
$F_{AB}$ to form a new Fisher matrix, where $\delta$ is the Kronecker
delta and $\sigma_{A}$ is the standard deviation of $\theta_{A}$
obtained through pre-existing observations. We limit our observables
to $X$=\{$H$, $D_{A}$, $I'_{\nu',{\rm dip}}/I_{\nu'}$\}
where $H$ is the Hubble coefficient, $D_{A}$ is the angular diameter
distance, and $I'_{\nu',{\rm dip}}/I_{\nu'}$ is the
ratio of the dipole to the monopole of a LIM. We only consider (1)
the combination of all of $H$, $D_{A}$ and LIM and (2) the LIM-only
observation but with priors on cosmological parameters.

\subsubsection{Case 1: Combined observation of $H$, $D_{A}$ and $I'_{\nu',{\rm dip}}/I_{\nu'}$}\label{subsec:combined}

We present our forecast on a combined observation of $H$, $D_{A}$
and $I'_{\nu',{\rm dip}}/I_{\nu'}$. Measuring $H$
and $D_{A}$ allows estimation of cosmological parameters. We assume
an almost ideal situation, where we have only 1\% error\footnote{We assume such a high accuracy on measures of $H$ and $D_{A}$ in order to achieve a marginalized posterior on $w_0$ comparable to the accuracy reported by a recent Alcock-Paczynski test \cite{Dong2023}. This test uses seemingly anisotropic clustering of galaxies and thus it is not straightforward to translate its accuracy to the measures of $H$ and $D_{A}$. Instead, we tried several values of the observational accuracy in $H$ and $D_{A}$ and found that the quoted accuracy leads to $\sim 9$\% error on the marginalized estimation of $w_0$. One may accept this choice as a reflection of the combined power of various comsmological observations.} on both $H$
and $D_{A}$ at each of the 20 redshift bins equally spaced at $z$={[}0,
2{]}, and 3\% error on $I'_{\nu',{\rm dip}}/I_{\nu'}$
at each of the 40 redshift bins equally spaced at $z$={[}0, 4{]}.
The reason to stretch the redshift range of $I'_{\nu',{\rm dip}}/I_{\nu'}$
is to include both the increasing and decreasing trends in $\rho_{{\rm c}}$
at $z>z_{{\rm t}}\simeq2$ and $z<z_{{\rm t}}\simeq2$, respectively.
If the redshift range of $z\gtrsim2$ were excluded but $\rho_{{\rm c}}$
followed the form of Eq. (\ref{eq:rhostar}), then $\gamma_{2}$
would not be constrained.

\begin{figure}[h]
\centering
\includegraphics[width=0.95\textwidth]{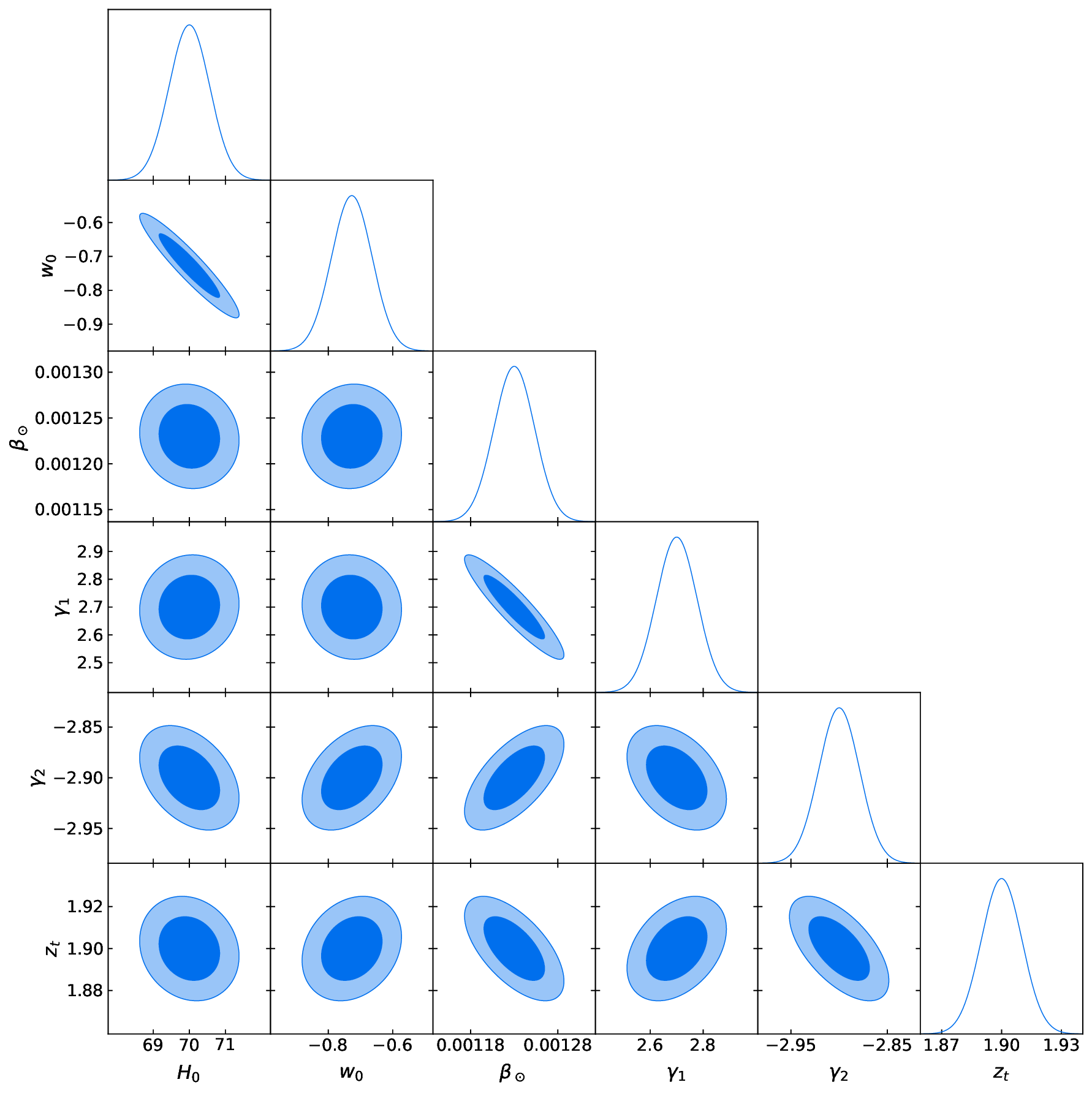}
\caption{Forecast for an ideal $H$+$D_{A}+I'_{\nu',{\rm dip}}/I_{\nu'}$
observation, assuming 1\% error on $H$ and $D_{A}$ at each of 40
redshift bins from z=0 to 2 and 3\% error on $I'_{\nu',{\rm dip}}/I_{\nu'}$
at each of 40 redshift bins from z=0 to 4. Expected estimation of
parameters $H_{0}$, $w_{0}$, $\beta_{\odot},$ $\gamma_{1}$, $\gamma_{2}$,
and $z_{{\rm t}}$ after marginalization over other cosmological parameters.
Shown are 2D contours of $1\sigma$ and $2\sigma$ confidence levels,
and the 1D posterior distributions.}\label{fig:combinedFC}
\end{figure}

Having such an accurate measurement of $H$ and $D_{A}$ would allow
a reliable estimation of not only the cosmological parameters but
also $\beta_{\odot}$, $\gamma_{1}$, $\gamma_{2}$, and $z_{{\rm {\rm t}}}$.
In the light of the recent measures of the baryon acoustic oscillation
(BAO) \cite{DESI2024} and an Alcock-Paczynski test \cite{Dong2023} hinting at non-standard cosmology
with $w_{0}\ne1$ and $w_{a}\ne0$, we just include these parameters
to have a triangle plot on marginalized posterior distributions in
Fig. \ref{fig:combinedFC}. Through the help of accurate measures
of $H$ and $D_{A}$, our Fisher matrix analysis results in the marginalized
1D estimations shown in Table \ref{tab:sigma}. Marginalized 2D posterior contours
of $1\sigma$ and $2\sigma$ confidence levels are shown in Fig.~\ref{fig:combinedFC}.

\begin{table}[h]
\caption{Parameter-estimation potential of two observational strategies (Sects.~\ref{subsec:combined} and \ref{subsec:dipole-only}), expressed in terms of per-cent error in each parameter.}\label{tab:sigma}%
\begin{tabular}{@{}lllllll@{}}
\toprule
parameters $\theta$ & $w_{0}$ & $w_{a}$ & $\beta_{\odot}$ & $\gamma_{1}$ & $\gamma_{2}$ & $z_{{\rm t}}$\tabularnewline
\midrule
fiducial values & -0.727 & -1.05 & 0.00123 & 2.7 & -2.9 & 1.9\tabularnewline
$1\sigma$ \% error for $H$+$D_{A}+I'_{\nu',{\rm dip}}/I_{\nu'}$ & 8.7 & 35 & 1.9 & 2.8 & 0.7 & 0.5\tabularnewline
$1\sigma$ \% error for $I'_{\nu',{\rm dip}}/I_{\nu'}$+(cosmological
priors) & (9.2)\footnotemark[1] & (28) & 9.7 & 12 & 2.5 & 2.8\tabularnewline
\botrule
\end{tabular}
\footnotetext{The per-cent errors are based on the fiducial values listed.}
\footnotetext[1]{Numbers in parentheses are priors.}
\end{table}

% \begin{table}
% \begin{tabular}{|c|c|c|c|c|c|c|}
% \hline 
% parameters $\theta$ & $w_{0}$ & $w_{a}$ & $\beta_{\odot}$ & $\gamma_{1}$ & $\gamma_{2}$ & $z_{{\rm t}}$\tabularnewline
% \hline 
% \hline 
% fiducial values & -0.727 & -1.05 & 0.00123 & 2.7 & -2.9 & 1.9\tabularnewline
% \hline 
% $1\sigma$ \% error for $H$+$D_{A}+\left.I'_{\nu'}\right)_{{\rm dip}}/I_{\nu'}$ & 8.7 & 35 & 1.9 & 2.8 & 0.7 & 0.5\tabularnewline
% \hline 
% $1\sigma$ \% error for $\left.I'_{\nu'}\right)_{{\rm dip}}/I_{\nu'}$+(cosmological
% priors) & (9.2) & (28) & 9.7 & 12 & 2.5 & 2.8\tabularnewline
% \hline 
% \end{tabular}

% \caption{Parameter-estimation potential of two observational strategies (Sections
% \ref{subsec:combined} and \ref{subsec:dipole-only}), expressed in
% terms of per-cent error in each parameter. Numbers in parentheses
% are priors. The per-cent error is based on the fiducial values listed.\label{tab:sigma}}

% \end{table}

\subsubsection{Case 2: Dipole-only measurement with priors on cosmological parameters}\label{subsec:dipole-only}

We now forecast a fractional dipole measurement of LIM, $I'_{\nu',{\rm dip}}/I_{\nu'}$,
combined with priors on cosmological parameters from pre-existing
observations. The priors we use here are $H_{0}=(70\pm1.1)\,{\rm km\,s^{-1}\,Mpc^{-1}}$,
$\Omega_{m,0}=0.3\pm0.023$, $\Omega_{k,0}=0\pm0.002$ \cite{PlanckCollaboration2018},
$w_{0}=-0.727\pm0.67$, and $w_{a}=-1.05\pm0.29$ \cite{DESI2024}. We do not here
encompass the ''Hubble tension'' between the estimations from the
CMB observation and the measurement of relatively local standard candles,
but instead just adopt the estimation from the latter. We then assume
an observation of $I'_{\nu',{\rm dip}}/I_{\nu'}$ that
has the observational error of $10\%$ at each of 20 uniformly spaced
redshift bins at $z$={[}0, 4{]}.

\begin{figure}[h]
\centering
\includegraphics[width=0.7\textwidth]{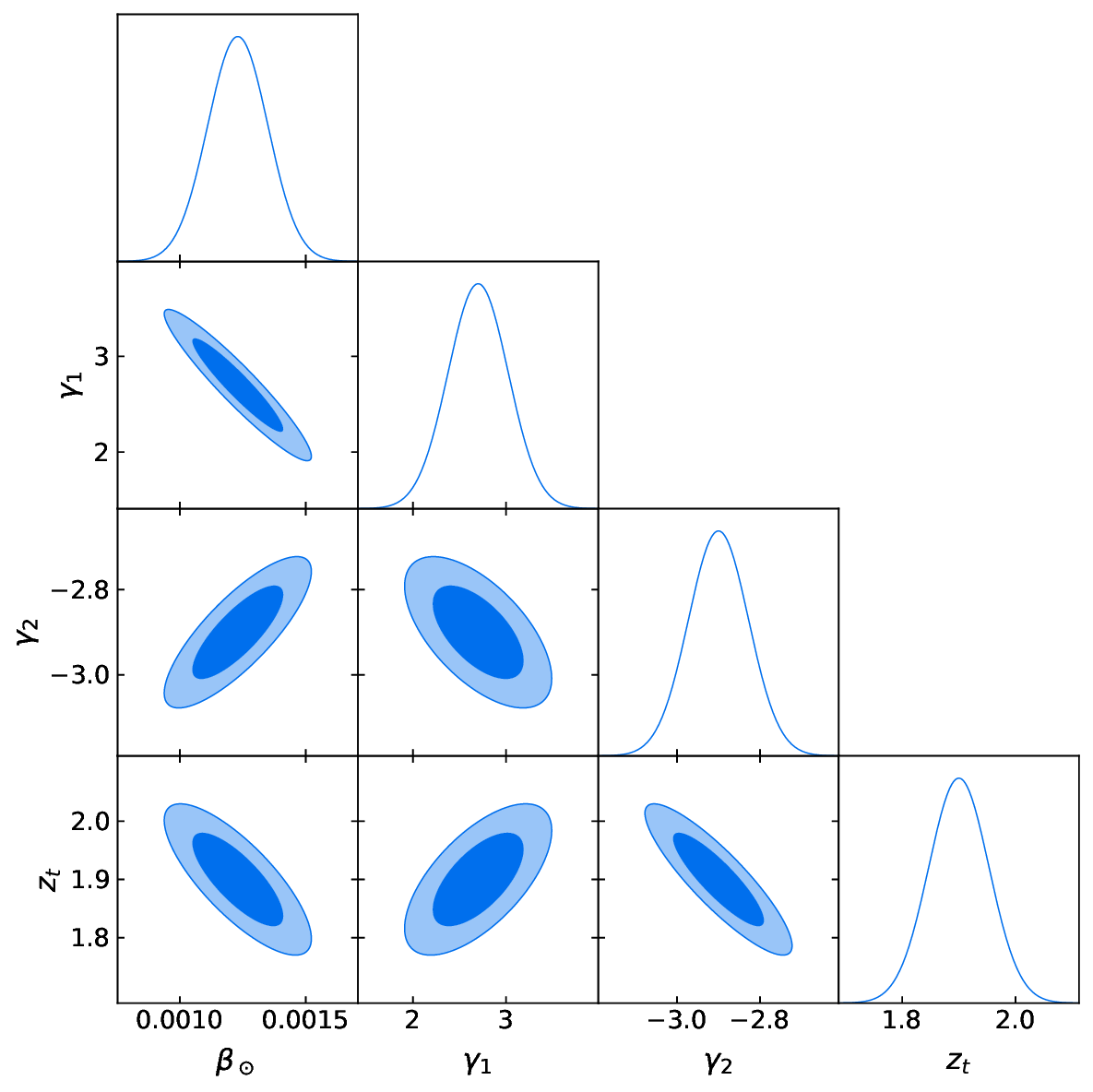}
\caption{Forecast for an $I'_{\nu',{\rm dip}}/I_{\nu'}$-only
observation at 10\% error at each of 40 redshift bins from z=0 to
4 with priors on cosmological parameters.}\label{fig:dipoleonly}
\end{figure}

The result is still promising as shown in Table~\ref{tab:sigma}
and Fig.~\ref{fig:dipoleonly}, even though the error budgets increase
to $\sim$5 times those of the combined observation (Sect. \ref{subsec:combined}).
All parameters are found to be with estimation error $\lesssim10\%$,
and especially $\beta_{\odot}$ with $9.7\%$ error. Table \ref{tab:sigma}
and Fig. \ref{fig:dipoleonly} shows the result. This result shows
that the dipole-only measurement can constrain both $\beta_{\odot}$
and the temporal evolution of $\rho_{{\rm c}}$, thus providing a
valuable information on astrophysics.

\section{Summary and Discussion}\label{sec:Discussion}

Spectral-line based mapping of galaxies or diffuse intensities can
reveal the dipole anisotropy due to the motion of our solar system.
These SGS and LIM dipoles are separated in frequencies, or equivalently
in redshifts of origin, and thus provide natural redundancy in the
estimation of the dipole. Therefore, we expect that such surveys could
provide a reliable constraint on $\beta_{\odot}$. Full-sky SGS such
as the upcoming SPHEREx allows a luxurious number of galaxies large
enough to provide an excellent constraint on $\beta_{\odot}$. As
for LIM, there does not exist a full-sky LIM yet proposed. There exist
some existing or planned telescopes with the spectroscopic capability
and the sky coverage of about $50-70\%$, e.g. 21-cm line survey instrument
SKA-LOW, which may be used
for the LIM dipole measurement if a pattern-matching scheme is adopted
to overcome the limitation of the incomplete sky coverage \cite{Secrest2021}.
In this work, we simulated the usage of a future full-sky LIM survey
data. Such an observation, combined with already-exiting prior constraints
on cosmological parameters, is found to be able to place strong constraints
on astrophysical parameters as well as $\beta_{\odot}$.
Such surveys would not require a high angular resolution,
and thus we propose a low angular resolution LIM survey that would
observe the sky with a high speed. We defer setting up a more detailed
observational strategy to a future paper.

We now discuss, for serendipity, a fictitious case where the estimated
$\beta_{\odot}$ from either the SGS dipole or the LIM dipole is found very
different from the CMB-dipole based constraint, $\beta_{\odot,{\rm CMB}}=(1.23\pm0.017)\times10^{-3}$.
Such a result can further be categorized into (1) the one with the
directional consistency and (2) the one with the directional inconsistency.
First, if there exists directional consistency, then there seems to
be only two possibilities: (a) the galaxy bias is strong enough to
contribute an additional dipole moment but galaxies still probe the
local structure that causes the motion of the solar system against
the CMB-rest frame, or (b) there exists a systematic uncertainty in
observation. Second, if there exists directional inconsistency, the
possibilities are (a) CMB-rest and matter-rest frames are moving against
each other, or (b) there again exists a systematic uncertainty in observation.
It is easy to blame the systematic uncertainties, but the evidences
for strong discrepancies seem to be piling up.

What if our local universe within $\lesssim3\,{\rm Gpc}$ were statistically
peculiar? There exist such symptoms indeed, and most notable one is
the discrepancy between the measures of the baryon acoustic oscillation
signature in the northern and southern Galactic caps \cite{Tojeiro2014}.
Even though this fact is usually ignored and attributed again to observation-based
uncertainties, it is interesting to take it as a true peculiarity
of our local universe and find its possible link to the dipole discrepancy.
The most natural way to link this north-south tension would be to
have the local matter-rest frame moving against the CMB-rest frame,
or simply having an additional effect on the dipole from the local
inhomogeneity of matter distribution. Now, the launch of SPHEREx is
imminent and thus the dipole anisotropy of the SGS by SPHEREx, in
a number large enough to provide us with a tight constraint on $\beta_{\odot}$,
is thus highly anticipated and would either deepen or simply resolve
the puzzle. We also argue that it is good time to seriously consider
full-sky LIM surveys for the same reason. Such surveys will provide
a deep insight into the present day cosmology that tends to lean toward
a possible non-standard cosmology other than the $\Lambda$CDM model.

\bmhead{Acknowledgements}

KA is supported by NRF-2021R1A2C1095136 and a research grant from Chosun University (2017). KA also acknowledges the hospitality of Curtin University, where part of this work was conducted.

%%\bibliography{../refs1}

\begin{thebibliography}{36}
\ifx \bisbn   \undefined \def \bisbn  #1{ISBN #1}\fi
\ifx \binits  \undefined \def \binits#1{#1}\fi
\ifx \bauthor  \undefined \def \bauthor#1{#1}\fi
\ifx \batitle  \undefined \def \batitle#1{#1}\fi
\ifx \bjtitle  \undefined \def \bjtitle#1{#1}\fi
\ifx \bvolume  \undefined \def \bvolume#1{\textbf{#1}}\fi
\ifx \byear  \undefined \def \byear#1{#1}\fi
\ifx \bissue  \undefined \def \bissue#1{#1}\fi
\ifx \bfpage  \undefined \def \bfpage#1{#1}\fi
\ifx \blpage  \undefined \def \blpage #1{#1}\fi
\ifx \burl  \undefined \def \burl#1{\textsf{#1}}\fi
\ifx \doiurl  \undefined \def \doiurl#1{\url{https://doi.org/#1}}\fi
\ifx \betal  \undefined \def \betal{\textit{et al.}}\fi
\ifx \binstitute  \undefined \def \binstitute#1{#1}\fi
\ifx \binstitutionaled  \undefined \def \binstitutionaled#1{#1}\fi
\ifx \bctitle  \undefined \def \bctitle#1{#1}\fi
\ifx \beditor  \undefined \def \beditor#1{#1}\fi
\ifx \bpublisher  \undefined \def \bpublisher#1{#1}\fi
\ifx \bbtitle  \undefined \def \bbtitle#1{#1}\fi
\ifx \bedition  \undefined \def \bedition#1{#1}\fi
\ifx \bseriesno  \undefined \def \bseriesno#1{#1}\fi
\ifx \blocation  \undefined \def \blocation#1{#1}\fi
\ifx \bsertitle  \undefined \def \bsertitle#1{#1}\fi
\ifx \bsnm \undefined \def \bsnm#1{#1}\fi
\ifx \bsuffix \undefined \def \bsuffix#1{#1}\fi
\ifx \bparticle \undefined \def \bparticle#1{#1}\fi
\ifx \barticle \undefined \def \barticle#1{#1}\fi
\bibcommenthead
\ifx \bconfdate \undefined \def \bconfdate #1{#1}\fi
\ifx \botherref \undefined \def \botherref #1{#1}\fi
\ifx \url \undefined \def \url#1{\textsf{#1}}\fi
\ifx \bchapter \undefined \def \bchapter#1{#1}\fi
\ifx \bbook \undefined \def \bbook#1{#1}\fi
\ifx \bcomment \undefined \def \bcomment#1{#1}\fi
\ifx \oauthor \undefined \def \oauthor#1{#1}\fi
\ifx \citeauthoryear \undefined \def \citeauthoryear#1{#1}\fi
\ifx \endbibitem  \undefined \def \endbibitem {}\fi
\ifx \bconflocation  \undefined \def \bconflocation#1{#1}\fi
\ifx \arxivurl  \undefined \def \arxivurl#1{\textsf{#1}}\fi
\csname PreBibitemsHook\endcsname

%%% 1
\bibitem[\protect\citeauthoryear{Yasini and Pierpaoli}{2017}]{Yasini2017}
\begin{barticle}
\bauthor{\bsnm{Yasini}, \binits{S.}},
\bauthor{\bsnm{Pierpaoli}, \binits{E.}}:
\batitle{Beyond the boost: Measuring the intrinsic dipole of the cosmic
  microwave background using the spectral distortions of the monopole and
  quadrupole}.
\bjtitle{\prl}
\bvolume{119}(\bissue{22}),
\bfpage{221102}
(\byear{2017})
\doiurl{10.1103/PhysRevLett.119.221102}
%{\href{https://arxiv.org/abs/1610.00015}{{arXiv:1610.00015}}}
%{[astro-ph.CO]}
\end{barticle}
\endbibitem

%%% 2
\bibitem[\protect\citeauthoryear{Ellis and Baldwin}{1984}]{Ellis1984}
\begin{barticle}
\bauthor{\bsnm{Ellis}, \binits{G.F.R.}},
\bauthor{\bsnm{Baldwin}, \binits{J.E.}}:
\batitle{On the expected anisotropy of radio source counts}.
\bjtitle{\mnras}
\bvolume{206},
\bfpage{377}--\blpage{381}
(\byear{1984})
\doiurl{10.1093/mnras/206.2.377}
\end{barticle}
\endbibitem

%%% 3
\bibitem[\protect\citeauthoryear{{Xia} et~al.}{2010}]{Xia2010}
\begin{barticle}
\bauthor{\bsnm{{Xia}}, \binits{J.-Q.}},
\bauthor{\bsnm{{Viel}}, \binits{M.}},
\bauthor{\bsnm{{Baccigalupi}}, \binits{C.}},
\bauthor{\bsnm{{De Zotti}}, \binits{G.}},
\bauthor{\bsnm{{Matarrese}}, \binits{S.}},
\bauthor{\bsnm{{Verde}}, \binits{L.}}:
\batitle{{Primordial Non-Gaussianity and the NRAO VLA Sky Survey}}.
\bjtitle{\apjl}
\bvolume{717}(\bissue{1}),
\bfpage{17}--\blpage{21}
(\byear{2010})
\doiurl{10.1088/2041-8205/717/1/L17}
%{\href{https://arxiv.org/abs/1003.3451}{{arXiv:1003.3451}}}
%{[astro-ph.CO]}
\end{barticle}
\endbibitem

%%% 4
\bibitem[\protect\citeauthoryear{Singal}{2011}]{Singal2011}
\begin{barticle}
\bauthor{\bsnm{Singal}, \binits{A.K.}}:
\batitle{Large peculiar motion of the solar system from the dipole anisotropy
  in sky brightness due to distant radio sources}.
\bjtitle{\apj}
\bvolume{742}(\bissue{2}),
\bfpage{23}
(\byear{2011})
\doiurl{10.1088/2041-8205/742/2/l23}
\end{barticle}
\endbibitem

%%% 5
\bibitem[\protect\citeauthoryear{{Chen} and {Schwarz}}{2016}]{Chen2016}
\begin{barticle}
\bauthor{\bsnm{{Chen}}, \binits{S.}},
\bauthor{\bsnm{{Schwarz}}, \binits{D.J.}}:
\batitle{{Angular two-point correlation of NVSS galaxies revisited}}.
\bjtitle{\aap}
\bvolume{591},
\bfpage{135}
(\byear{2016})
\doiurl{10.1051/0004-6361/201526956}
%{\href{https://arxiv.org/abs/1507.02160}{{arXiv:1507.02160}}}
%{[astro-ph.CO]}
\end{barticle}
\endbibitem

%%% 6
\bibitem[\protect\citeauthoryear{{Secrest} et~al.}{2021}]{Secrest2021}
\begin{barticle}
\bauthor{\bsnm{{Secrest}}, \binits{N.J.}},
\bauthor{\bsnm{{von Hausegger}}, \binits{S.}},
\bauthor{\bsnm{{Rameez}}, \binits{M.}},
\bauthor{\bsnm{{Mohayaee}}, \binits{R.}},
\bauthor{\bsnm{{Sarkar}}, \binits{S.}},
\bauthor{\bsnm{{Colin}}, \binits{J.}}:
\batitle{{A Test of the Cosmological Principle with Quasars}}.
\bjtitle{\apjl}
\bvolume{908}(\bissue{2}),
\bfpage{51}
(\byear{2021})
\doiurl{10.3847/2041-8213/abdd40}
%{\href{https://arxiv.org/abs/2009.14826}{{arXiv:2009.14826}}}
%{[astro-ph.CO]}
\end{barticle}
\endbibitem

%%% 7
\bibitem[\protect\citeauthoryear{Dalang and Bonvin}{2022}]{Dalang2022}
\begin{barticle}
\bauthor{\bsnm{Dalang}, \binits{C.}},
\bauthor{\bsnm{Bonvin}, \binits{C.}}:
\batitle{On the kinematic cosmic dipole tension}.
\bjtitle{\mnras}
\bvolume{512}(\bissue{3}),
\bfpage{3895}--\blpage{3905}
(\byear{2022})
\doiurl{10.1093/mnras/stac726}
%{\href{https://arxiv.org/abs/2111.03616}{{arXiv:2111.03616}}}
%{[astro-ph.CO]}
\end{barticle}
\endbibitem

%%% 8
\bibitem[\protect\citeauthoryear{Kumar~Aluri et~al.}{2023}]{KumarAluri2023}
\begin{barticle}
\bauthor{\bsnm{Kumar~Aluri}, \binits{P.}},
\bauthor{\bsnm{Cea}, \binits{P.}},
\bauthor{\bsnm{Chingangbam}, \binits{P.}},
\bauthor{\bsnm{Chu}, \binits{M.-C.}},
\bauthor{\bsnm{Clowes}, \binits{R.G.}},
\bauthor{\bsnm{Hutsem{\'e}kers}, \binits{D.}},
\bauthor{\bsnm{Kochappan}, \binits{J.P.}},
\bauthor{\bsnm{Lopez}, \binits{A.M.}},
\bauthor{\bsnm{Liu}, \binits{L.}},
\bauthor{\bsnm{Martens}, \binits{N.C.M. et al.}}:
% \bauthor{\bsnm{Martins}, \binits{C.J.A.P.}},
% \bauthor{\bsnm{Migkas}, \binits{K.}},
% \bauthor{\bsnm{\'O~Colg{\'a}in}, \binits{E.}},
% \bauthor{\bsnm{Pranav}, \binits{P.}},
% \bauthor{\bsnm{Shamir}, \binits{L.}},
% \bauthor{\bsnm{Singal}, \binits{A.K.}},
% \bauthor{\bsnm{Sheikh-Jabbari}, \binits{M.M.}},
% \bauthor{\bsnm{Wagner}, \binits{J.}},
% \bauthor{\bsnm{Wang}, \binits{S.-J.}},
% \bauthor{\bsnm{Wiltshire}, \binits{D.L.}},
% \bauthor{\bsnm{Yeung}, \binits{S.}},
% \bauthor{\bsnm{Yin}, \binits{L.}},
% \bauthor{\bsnm{Zhao}, \binits{W.}}:
\batitle{Is the observable universe consistent with the cosmological
  principle?}
\bjtitle{\cqg}
\bvolume{40}(\bissue{9}),
\bfpage{094001}
(\byear{2023})
\doiurl{10.1088/1361-6382/acbefc}
%{\href{https://arxiv.org/abs/2207.05765}{{arXiv:2207.05765}}}
%{[astro-ph.CO]}
\end{barticle}
\endbibitem

%%% 9
\bibitem[\protect\citeauthoryear{Darling}{2022}]{Darling2022}
\begin{barticle}
\bauthor{\bsnm{Darling}, \binits{J.}}:
\batitle{The universe is brighter in the direction of our motion: Galaxy counts
  and fluxes are consistent with the cmb dipole}.
\bjtitle{\apjl}
\bvolume{931}(\bissue{2}),
\bfpage{14}
(\byear{2022})
\doiurl{10.3847/2041-8213/ac6f08}
%{\href{https://arxiv.org/abs/2205.06880}{{arXiv:2205.06880}}}
%{[astro-ph.CO]}
\end{barticle}
\endbibitem

%%% 10
\bibitem[\protect\citeauthoryear{Panwar et~al.}{2024}]{Panwar2024}
\begin{barticle}
\bauthor{\bsnm{Panwar}, \binits{M.}},
\bauthor{\bsnm{Jain}, \binits{P.}},
\bauthor{\bsnm{Omar}, \binits{A.}}:
\batitle{Colour dependence of dipole in catwise2020 data}.
\bjtitle{\mnrasl}
\bvolume{535}(\bissue{1}),
\bfpage{63}--\blpage{69}
(\byear{2024})
\doiurl{10.1093/mnrasl/slae093}
\end{barticle}
\endbibitem

%%% 11
\bibitem[\protect\citeauthoryear{Tiwari and Nusser}{2016}]{Tiwari2016}
\begin{barticle}
\bauthor{\bsnm{Tiwari}, \binits{P.}},
\bauthor{\bsnm{Nusser}, \binits{A.}}:
\batitle{Revisiting the nvss number count dipole}.
\bjtitle{\jcap}
\bvolume{2016}(\bissue{03}),
\bfpage{062}--\blpage{062}
(\byear{2016})
\doiurl{10.1088/1475-7516/2016/03/062}
\end{barticle}
\endbibitem

%%% 12
\bibitem[\protect\citeauthoryear{Domènech et~al.}{2022}]{Domenech2022}
\begin{barticle}
\bauthor{\bsnm{Domènech}, \binits{G.}},
\bauthor{\bsnm{Mohayaee}, \binits{R.}},
\bauthor{\bsnm{Patil}, \binits{S.P.}},
\bauthor{\bsnm{Sarkar}, \binits{S.}}:
\batitle{Galaxy number-count dipole and superhorizon fluctuations}.
\bjtitle{Journal of Cosmology and Astroparticle Physics}
\bvolume{2022}(\bissue{10}),
\bfpage{019}
(\byear{2022})
\doiurl{10.1088/1475-7516/2022/10/019}
\end{barticle}
\endbibitem

%%% 13
\bibitem[\protect\citeauthoryear{Crill et~al.}{2020}]{Crill2020}
\begin{bchapter}
\bauthor{\bsnm{Crill}, \binits{B.P.}},
\bauthor{\bsnm{Werner}, \binits{M.}},
\bauthor{\bsnm{Akeson}, \binits{R.}},
\bauthor{\bsnm{Ashby}, \binits{M.}},
\bauthor{\bsnm{Bleem}, \binits{L.}},
\bauthor{\bsnm{Bock}, \binits{J.J.}},
\bauthor{\bsnm{Bryan}, \binits{S.}},
\bauthor{\bsnm{Burnham}, \binits{J.}},
\bauthor{\bsnm{Byunh}, \binits{J.}},
\bauthor{\bsnm{Chang}, \binits{T.-C. et~al.}}:
% \bauthor{\bsnm{Chiang}, \binits{Y.-K.}},
% \bauthor{\bsnm{Cook}, \binits{W.}},
% \bauthor{\bsnm{Cooray}, \binits{A.}},
% \bauthor{\bsnm{Davis}, \binits{A.}},
% \bauthor{\bsnm{Dor{\'e}}, \binits{O.}},
% \bauthor{\bsnm{Dowell}, \binits{C.D.}},
% \bauthor{\bsnm{Dubois-Felsmann}, \binits{G.}},
% \bauthor{\bsnm{Eifler}, \binits{T.}},
% \bauthor{\bsnm{Faisst}, \binits{A.}},
% \bauthor{\bsnm{Habib}, \binits{S.}},
% \bauthor{\bsnm{Heinrich}, \binits{C.}},
% \bauthor{\bsnm{Heitmann}, \binits{K.}},
% \bauthor{\bsnm{Heaton}, \binits{G.}},
% \bauthor{\bsnm{Hirata}, \binits{C.}},
% \bauthor{\bsnm{Hristov}, \binits{V.}},
% \bauthor{\bsnm{Hui}, \binits{H.}},
% \bauthor{\bsnm{Jeong}, \binits{W.-S.}},
% \bauthor{\bsnm{Kang}, \binits{J.H.}},
% \bauthor{\bsnm{Kecman}, \binits{B.}},
% \bauthor{\bsnm{Kirkpatrick}, \binits{J.D.}},
% \bauthor{\bsnm{Korngut}, \binits{P.M.}},
% \bauthor{\bsnm{Krause}, \binits{E.}},
% \bauthor{\bsnm{Lee}, \binits{B.}},
% \bauthor{\bsnm{Lisse}, \binits{C.}},
% \bauthor{\bsnm{Masters}, \binits{D.}},
% \bauthor{\bsnm{Mauskopf}, \binits{P.}},
% \bauthor{\bsnm{Melnick}, \binits{G.}},
% \bauthor{\bsnm{Miyasaka}, \binits{H.}},
% \bauthor{\bsnm{Nayyeri}, \binits{H.}},
% \bauthor{\bsnm{Nguyen}, \binits{H.}},
% \bauthor{\bsnm{{\"O}berg}, \binits{K.}},
% \bauthor{\bsnm{Padin}, \binits{S.}},
% \bauthor{\bsnm{Paladini}, \binits{R.}},
% \bauthor{\bsnm{Pourrahmani}, \binits{M.}},
% \bauthor{\bsnm{Pyo}, \binits{J.}},
% \bauthor{\bsnm{Smith}, \binits{R.}},
% \bauthor{\bsnm{Song}, \binits{Y.-S.}},
% \bauthor{\bsnm{Symons}, \binits{T.}},
% \bauthor{\bsnm{Teplitz}, \binits{H.}},
% \bauthor{\bsnm{Tolls}, \binits{V.}},
% \bauthor{\bsnm{Unwin}, \binits{S.}},
% \bauthor{\bsnm{Windhorst}, \binits{R.}},
% \bauthor{\bsnm{Yang}, \binits{Y.}},
% \bauthor{\bsnm{Zemcov}, \binits{M.}}:
\bctitle{Spherex: Nasa's near-infrared spectrophotometric all-sky survey}.
In: \beditor{\bsnm{{Lystrup}}, \binits{M.}},
\beditor{\bsnm{{Perrin}}, \binits{M.D.}} (eds.)
\bbtitle{Space Telescopes and Instrumentation 2020: Optical, Infrared, and
  Millimeter Wave}.
\bsertitle{Society of Photo-Optical Instrumentation Engineers (SPIE) Conference
  Series},
vol. \bseriesno{11443},
p. \bfpage{114430}
(\byear{2020}).
\doiurl{10.1117/12.2567224} .
%\burl{https://ui.adsabs.harvard.edu/abs/2020SPIE11443E..0IC}
\end{bchapter}
\endbibitem

%%% 14
\bibitem[\protect\citeauthoryear{Mertens et~al.}{2020}]{Mertens2020}
\begin{barticle}
\bauthor{\bsnm{Mertens}, \binits{F.G.}},
\bauthor{\bsnm{Mevius}, \binits{M.}},
\bauthor{\bsnm{Koopmans}, \binits{L.V.E.}},
\bauthor{\bsnm{Offringa}, \binits{A.R.}},
\bauthor{\bsnm{Mellema}, \binits{G.}},
\bauthor{\bsnm{Zaroubi}, \binits{S.}},
\bauthor{\bsnm{Brentjens}, \binits{M.A.}},
\bauthor{\bsnm{Gan}, \binits{H.}},
\bauthor{\bsnm{Gehlot}, \binits{B.K.}},
\bauthor{\bsnm{Pand~ey}, \binits{V.N. et al.}}:
% \bauthor{\bsnm{Sardarabadi}, \binits{A.M.}},
% \bauthor{\bsnm{Vedantham}, \binits{H.K.}},
% \bauthor{\bsnm{Yatawatta}, \binits{S.}},
% \bauthor{\bsnm{Asad}, \binits{K.M.B.}},
% \bauthor{\bsnm{Ciardi}, \binits{B.}},
% \bauthor{\bsnm{Chapman}, \binits{E.}},
% \bauthor{\bsnm{Gazagnes}, \binits{S.}},
% \bauthor{\bsnm{Ghara}, \binits{R.}},
% \bauthor{\bsnm{Ghosh}, \binits{A.}},
% \bauthor{\bsnm{Giri}, \binits{S.K.}},
% \bauthor{\bsnm{Iliev}, \binits{I.T.}},
% \bauthor{\bsnm{Jeli{\'c}}, \binits{V.}},
% \bauthor{\bsnm{Kooistra}, \binits{R.}},
% \bauthor{\bsnm{Mondal}, \binits{R.}},
% \bauthor{\bsnm{Schaye}, \binits{J.}},
% \bauthor{\bsnm{Silva}, \binits{M.B.}}:
\batitle{Improved upper limits on the 21 cm signal power spectrum of neutral
  hydrogen at z {\ensuremath{\approx}} 9.1 from lofar}.
\bjtitle{\mnras}
\bvolume{493}(\bissue{2}),
\bfpage{1662}--\blpage{1685}
(\byear{2020})
\doiurl{10.1093/mnras/staa327}
%{\href{https://arxiv.org/abs/2002.07196}{{arXiv:2002.07196}}}
%{[astro-ph.CO]}
\end{barticle}
\endbibitem

%%% 15
\bibitem[\protect\citeauthoryear{Ahn et~al.}{2015}]{Ahn2015b}
\begin{bchapter}
\bauthor{\bsnm{Ahn}, \binits{K.}},
\bauthor{\bsnm{Mesinger}, \binits{A.}},
\bauthor{\bsnm{Alvarez}, \binits{M.A.}},
\bauthor{\bsnm{Chen}, \binits{X.}}:
\bctitle{Probing the first galaxies and their impact on the intergalactic
  medium through 21-cm observations of the cosmic dawn with the ska}.
In: \bbtitle{Advancing Astrophysics with the Square Kilometre Array (AASKA14)},
p. \bfpage{3}
(\byear{2015}).
\doiurl{10.22323/1.215.0003} .
%\burl{https://ui.adsabs.harvard.edu/abs/2015aska.confE...3A}
\end{bchapter}
\endbibitem

%%% 16
\bibitem[\protect\citeauthoryear{Labate et~al.}{2022}]{Labate2022}
\begin{botherref}
\oauthor{\bsnm{Labate}, \binits{M.G.}},
\oauthor{\bsnm{Waterson}, \binits{M.}},
\oauthor{\bsnm{Alachkar}, \binits{B.}},
\oauthor{\bsnm{Hendre}, \binits{A.}},
\oauthor{\bsnm{Lewis}, \binits{P.}},
\oauthor{\bsnm{Bartolini}, \binits{M.}},
\oauthor{\bsnm{Dewdney}, \binits{P.}}:
Highlights of the square kilometre array low frequency (ska-low) telescope.
\bjtitle{\jatis}
\textbf{8}(01)
(2022)
\doiurl{10.1117/1.jatis.8.1.011024}
\end{botherref}
\endbibitem

%%% 17
\bibitem[\protect\citeauthoryear{Hotinli and Ahn}{2024}]{Hotinli2024}
\begin{barticle}
\bauthor{\bsnm{Hotinli}, \binits{S.C.}},
\bauthor{\bsnm{Ahn}, \binits{K.}}:
\batitle{Probing the global 21 cm background by velocity-induced dipole and
  quadrupole anisotropies}.
\bjtitle{\apj}
\bvolume{964}(\bissue{1}),
\bfpage{21}
(\byear{2024})
\doiurl{10.3847/1538-4357/ad2209}
%{\href{https://arxiv.org/abs/2305.01672}{{arXiv:2305.01672}}}
%{[astro-ph.CO]}
\end{barticle}
\endbibitem

%%% 18
\bibitem[\protect\citeauthoryear{Ahn and Oh}{2024}]{Ahn2024}
\begin{barticle}
\bauthor{\bsnm{Ahn}, \binits{K.}},
\bauthor{\bsnm{Oh}, \binits{M.}}:
\batitle{Probing the global 21-cm signal via the integrated sachs-wolfe effect
  on the 21-cm background}.
\bjtitle{\prd}
\bvolume{109}(\bissue{4}),
\bfpage{043539}
(\byear{2024})
\doiurl{10.1103/PhysRevD.109.043539}
%{\href{https://arxiv.org/abs/2308.14808}{{arXiv:2308.14808}}}
%{[astro-ph.CO]}
\end{barticle}
\endbibitem

%%% 19
\bibitem[\protect\citeauthoryear{Bottani et~al.}{1992}]{Bottani1992}
\begin{barticle}
\bauthor{\bsnm{Bottani}, \binits{S.}},
\bauthor{\bsnm{Bernardis}, \binits{P.}},
\bauthor{\bsnm{Melchiorri}, \binits{F.}}:
\batitle{On the origin of the dipole anisotropy as determined by quadrupole
  measurements}.
\bjtitle{\apjl}
\bvolume{384},
\bfpage{1}
(\byear{1992})
\doiurl{10.1086/186250}
\end{barticle}
\endbibitem

%%% 20
\bibitem[\protect\citeauthoryear{{Challinor} and {van
  Leeuwen}}{2002}]{Challinor2002}
\begin{barticle}
\bauthor{\bsnm{{Challinor}}, \binits{A.}},
\bauthor{\bsnm{{van Leeuwen}}, \binits{F.}}:
\batitle{{Peculiar velocity effects in high-resolution microwave background
  experiments}}.
\bjtitle{\prd}
\bvolume{65}(\bissue{10}),
\bfpage{103001}
(\byear{2002})
\doiurl{10.1103/PhysRevD.65.103001}
%{\href{https://arxiv.org/abs/astro-ph/0112457}{{arXiv:astro-ph/0112457}}}
%{[astro-ph]}
\end{barticle}
\endbibitem

%%% 21
\bibitem[\protect\citeauthoryear{{Rybicki} and
  {Lightman}}{1979}]{Rybicki_Lightman}
\begin{bbook}
\bauthor{\bsnm{{Rybicki}}, \binits{G.B.}},
\bauthor{\bsnm{{Lightman}}, \binits{A.P.}}:
\bbtitle{{Radiative Processes in Astrophysics}}
(\byear{1979})
\end{bbook}
\endbibitem

%%% 22
\bibitem[\protect\citeauthoryear{Peacock}{1999}]{Peacock1999}
\begin{bbook}
\bauthor{\bsnm{Peacock}, \binits{J.A.}}:
\bbtitle{Cosmological Physics}
(\byear{1999}).
%\burl{https://ui.adsabs.harvard.edu/abs/1999coph.book.....P}
\end{bbook}
\endbibitem

%%% 23
\bibitem[\protect\citeauthoryear{Madau and Dickinson}{2014}]{Madau2014}
\begin{barticle}
\bauthor{\bsnm{Madau}, \binits{P.}},
\bauthor{\bsnm{Dickinson}, \binits{M.}}:
\batitle{Cosmic star-formation history}.
\bjtitle{\araa}
\bvolume{52},
\bfpage{415}--\blpage{486}
(\byear{2014})
\doiurl{10.1146/annurev-astro-081811-125615}
%{\href{https://arxiv.org/abs/1403.0007}{{arXiv:1403.0007}}}
%{[astro-ph.CO]}
\end{barticle}
\endbibitem

%%% 24
\bibitem[\protect\citeauthoryear{{Chevallier} and
  {Polarski}}{2001}]{Chevallier2001}
\begin{barticle}
\bauthor{\bsnm{{Chevallier}}, \binits{M.}},
\bauthor{\bsnm{{Polarski}}, \binits{D.}}:
\batitle{{Accelerating Universes with Scaling Dark Matter}}.
\bjtitle{International Journal of Modern Physics D}
\bvolume{10}(\bissue{2}),
\bfpage{213}--\blpage{223}
(\byear{2001})
\doiurl{10.1142/S0218271801000822}
%{\href{https://arxiv.org/abs/gr-qc/0009008}{{arXiv:gr-qc/0009008}}}
%{[gr-qc]}
\end{barticle}
\endbibitem

%%% 25
\bibitem[\protect\citeauthoryear{{Linder}}{2003}]{Linder2003}
\begin{barticle}
\bauthor{\bsnm{{Linder}}, \binits{E.V.}}:
\batitle{{Exploring the Expansion History of the Universe}}.
\bjtitle{\prl}
\bvolume{90}(\bissue{9}),
\bfpage{091301}
(\byear{2003})
\doiurl{10.1103/PhysRevLett.90.091301}
%{\href{https://arxiv.org/abs/astro-ph/0208512}{{arXiv:astro-ph/0208512}}}
%{[astro-ph]}
\end{barticle}
\endbibitem

%%% 26
\bibitem[\protect\citeauthoryear{Abghari et~al.}{2024}]{Abghari2024}
\begin{botherref}
\oauthor{\bsnm{Abghari}, \binits{A.}},
\oauthor{\bsnm{Bunn}, \binits{E.F.}},
\oauthor{\bsnm{Hergt}, \binits{L.T.}},
\oauthor{\bsnm{Li}, \binits{B.}},
\oauthor{\bsnm{Scott}, \binits{D.}},
\oauthor{\bsnm{Sullivan}, \binits{R.M.}},
\oauthor{\bsnm{Wei}, \binits{D.}}:
Reassessment of the dipole in the distribution of quasars on the sky.
arXiv e-prints,
2405.09762
(2024)
\doiurl{10.48550/arXiv.2405.09762}
%{\href{https://arxiv.org/abs/2405.09762}{{arXiv:2405.09762}}}
%{[astro-ph.CO]}
\end{botherref}
\endbibitem

%%% 27
\bibitem[\protect\citeauthoryear{{\"O}stlin et~al.}{2014}]{Oestlin2014}
\begin{barticle}
\bauthor{\bsnm{{\"O}stlin}, \binits{G.}},
\bauthor{\bsnm{Hayes}, \binits{M.}},
\bauthor{\bsnm{Duval}, \binits{F.}},
\bauthor{\bsnm{Sandberg}, \binits{A.}},
\bauthor{\bsnm{Rivera-Thorsen}, \binits{T.}},
\bauthor{\bsnm{Marquart}, \binits{T.}},
\bauthor{\bsnm{Orlitov{\'a}}, \binits{I.}},
\bauthor{\bsnm{Adamo}, \binits{A.}},
\bauthor{\bsnm{Melinder}, \binits{J.}},
\bauthor{\bsnm{Guaita}, \binits{L. et al.}}:
\batitle{The ly{\ensuremath{\alpha}} reference sample. i. survey outline and
  first results for markarian 259}.
\bjtitle{\apj}
\bvolume{797}(\bissue{1}),
\bfpage{11}
(\byear{2014})
\doiurl{10.1088/0004-637X/797/1/11}
%{\href{https://arxiv.org/abs/1409.8347}{{arXiv:1409.8347}}}
%{[astro-ph.GA]}
\end{barticle}
\endbibitem

%%% 28
\bibitem[\protect\citeauthoryear{Ahumada et~al.}{2020}]{Ahumada2020}
\begin{barticle}
\bauthor{\bsnm{Ahumada}, \binits{R.}},
\bauthor{\bsnm{Allende~Prieto}, \binits{C.}},
\bauthor{\bsnm{Almeida}, \binits{A.}},
\bauthor{\bsnm{Anders}, \binits{F.}},
\bauthor{\bsnm{Anderson}, \binits{S.F.}},
\bauthor{\bsnm{Andrews}, \binits{B.H.}},
\bauthor{\bsnm{Anguiano}, \binits{B.}},
\bauthor{\bsnm{Arcodia}, \binits{R.}},
\bauthor{\bsnm{Armengaud}, \binits{E.}},
\bauthor{\bsnm{Aubert}, \binits{M. et al.}}:
\batitle{The 16th data release of the sloan digital sky surveys: First release
  from the apogee-2 southern survey and full release of eboss spectra}.
\bjtitle{\apjs}
\bvolume{249}(\bissue{1}),
\bfpage{3}
(\byear{2020})
\doiurl{10.3847/1538-4365/ab929e}
%{\href{https://arxiv.org/abs/1912.02905}{{arXiv:1912.02905}}}
%{[astro-ph.GA]}
\end{barticle}
\endbibitem

%%% 29
\bibitem[\protect\citeauthoryear{{HERA Collaboration} et~al.}{2023}]{HERA2023}
\begin{barticle}
\bauthor{\bsnm{{HERA Collaboration}}},
\bauthor{\bsnm{Abdurashidova}, \binits{Z.}},
\bauthor{\bsnm{Adams}, \binits{T.}},
\bauthor{\bsnm{Aguirre}, \binits{J.E.}},
\bauthor{\bsnm{Alexander}, \binits{P.}},
\bauthor{\bsnm{Ali}, \binits{Z.S.}},
\bauthor{\bsnm{Baartman}, \binits{R.}},
\bauthor{\bsnm{Balfour}, \binits{Y.}},
\bauthor{\bsnm{Barkana}, \binits{R.}},
\bauthor{\bsnm{Beardsley}, \binits{A.P. et al.}}:
% \bauthor{\bsnm{Bernardi}, \binits{G.}},
% \bauthor{\bsnm{Billings}, \binits{T.S.}},
% \bauthor{\bsnm{Bowman}, \binits{J.D.}},
% \bauthor{\bsnm{Bradley}, \binits{R.F.}},
% \bauthor{\bsnm{Breitman}, \binits{D.}},
% \bauthor{\bsnm{Bull}, \binits{P.}},
% \bauthor{\bsnm{Burba}, \binits{J.}},
% \bauthor{\bsnm{Carey}, \binits{S.}},
% \bauthor{\bsnm{Carilli}, \binits{C.L.}},
% \bauthor{\bsnm{Cheng}, \binits{C.}},
% \bauthor{\bsnm{Choudhuri}, \binits{S.}},
% \bauthor{\bsnm{DeBoer}, \binits{D.R.}},
% \bauthor{\bsnm{Lera~Acedo}, \binits{E.}},
% \bauthor{\bsnm{Dexter}, \binits{M.}},
% \bauthor{\bsnm{Dillon}, \binits{J.S.}},
% \bauthor{\bsnm{Ely}, \binits{J.}},
% \bauthor{\bsnm{Ewall-Wice}, \binits{A.}},
% \bauthor{\bsnm{Fagnoni}, \binits{N.}},
% \bauthor{\bsnm{Fialkov}, \binits{A.}},
% \bauthor{\bsnm{Fritz}, \binits{R.}},
% \bauthor{\bsnm{Furlanetto}, \binits{S.R.}},
% \bauthor{\bsnm{Gale-Sides}, \binits{K.}},
% \bauthor{\bsnm{Garsden}, \binits{H.}},
% \bauthor{\bsnm{Glendenning}, \binits{B.}},
% \bauthor{\bsnm{Gorce}, \binits{A.}},
% \bauthor{\bsnm{Gorthi}, \binits{D.}},
% \bauthor{\bsnm{Greig}, \binits{B.}},
% \bauthor{\bsnm{Grobbelaar}, \binits{J.}},
% \bauthor{\bsnm{Halday}, \binits{Z.}},
% \bauthor{\bsnm{Hazelton}, \binits{B.J.}},
% \bauthor{\bsnm{Heimersheim}, \binits{S.}},
% \bauthor{\bsnm{Hewitt}, \binits{J.N.}},
% \bauthor{\bsnm{Hickish}, \binits{J.}},
% \bauthor{\bsnm{Jacobs}, \binits{D.C.}},
% \bauthor{\bsnm{Julius}, \binits{A.}},
% \bauthor{\bsnm{Kern}, \binits{N.S.}},
% \bauthor{\bsnm{Kerrigan}, \binits{J.}},
% \bauthor{\bsnm{Kittiwisit}, \binits{P.}},
% \bauthor{\bsnm{Kohn}, \binits{S.A.}},
% \bauthor{\bsnm{Kolopanis}, \binits{M.}},
% \bauthor{\bsnm{Lanman}, \binits{A.}},
% \bauthor{\bsnm{La~Plante}, \binits{P.}},
% \bauthor{\bsnm{Lewis}, \binits{D.}},
% \bauthor{\bsnm{Liu}, \binits{A.}},
% \bauthor{\bsnm{Loots}, \binits{A.}},
% \bauthor{\bsnm{Ma}, \binits{Y.-Z.}},
% \bauthor{\bsnm{MacMahon}, \binits{D.H.E.}},
% \bauthor{\bsnm{Malan}, \binits{L.}},
% \bauthor{\bsnm{Malgas}, \binits{K.}},
% \bauthor{\bsnm{Malgas}, \binits{C.}},
% \bauthor{\bsnm{Maree}, \binits{M.}},
% \bauthor{\bsnm{Marero}, \binits{B.}},
% \bauthor{\bsnm{Martinot}, \binits{Z.E.}},
% \bauthor{\bsnm{McBride}, \binits{L.}},
% \bauthor{\bsnm{Mesinger}, \binits{A.}},
% \bauthor{\bsnm{Mirocha}, \binits{J.}},
% \bauthor{\bsnm{Molewa}, \binits{M.}},
% \bauthor{\bsnm{Morales}, \binits{M.F.}},
% \bauthor{\bsnm{Mosiane}, \binits{T.}},
% \bauthor{\bsnm{Mu{\~n}oz}, \binits{J.B.}},
% \bauthor{\bsnm{Murray}, \binits{S.G.}},
% \bauthor{\bsnm{Nagpal}, \binits{V.}},
% \bauthor{\bsnm{Neben}, \binits{A.R.}},
% \bauthor{\bsnm{Nikolic}, \binits{B.}},
% \bauthor{\bsnm{Nunhokee}, \binits{C.D.}},
% \bauthor{\bsnm{Nuwegeld}, \binits{H.}},
% \bauthor{\bsnm{Parsons}, \binits{A.R.}},
% \bauthor{\bsnm{Pascua}, \binits{R.}},
% \bauthor{\bsnm{Patra}, \binits{N.}},
% \bauthor{\bsnm{Pieterse}, \binits{S.}},
% \bauthor{\bsnm{Qin}, \binits{Y.}},
% \bauthor{\bsnm{Razavi-Ghods}, \binits{N.}},
% \bauthor{\bsnm{Robnett}, \binits{J.}},
% \bauthor{\bsnm{Rosie}, \binits{K.}},
% \bauthor{\bsnm{Santos}, \binits{M.G.}},
% \bauthor{\bsnm{Sims}, \binits{P.}},
% \bauthor{\bsnm{Singh}, \binits{S.}},
% \bauthor{\bsnm{Smith}, \binits{C.}},
% \bauthor{\bsnm{Swarts}, \binits{H.}},
% \bauthor{\bsnm{Tan}, \binits{J.}},
% \bauthor{\bsnm{Thyagarajan}, \binits{N.}},
% \bauthor{\bsnm{Wilensky}, \binits{M.J.}},
% \bauthor{\bsnm{Williams}, \binits{P.K.G.}},
% \bauthor{\bsnm{Wyngaarden}, \binits{P.}},
% \bauthor{\bsnm{Zheng}, \binits{H.}}:
\batitle{Improved constraints on the 21 cm eor power spectrum and the x-ray
  heating of the igm with hera phase i observations}.
\bjtitle{\apj}
\bvolume{945}(\bissue{2}),
\bfpage{124}
(\byear{2023})
\doiurl{10.3847/1538-4357/acaf50}
\end{barticle}
\endbibitem

%%% 30
\bibitem[\protect\citeauthoryear{Keating et~al.}{2016}]{Keating2016}
\begin{barticle}
\bauthor{\bsnm{Keating}, \binits{G.K.}},
\bauthor{\bsnm{Marrone}, \binits{D.P.}},
\bauthor{\bsnm{Bower}, \binits{G.C.}},
\bauthor{\bsnm{Leitch}, \binits{E.}},
\bauthor{\bsnm{Carlstrom}, \binits{J.E.}},
\bauthor{\bsnm{DeBoer}, \binits{D.R.}}:
\batitle{Copss ii: The molecular gas content of ten million cubic megaparsecs
  at redshift z {\ensuremath{\sim}} 3}.
\bjtitle{\apj}
\bvolume{830}(\bissue{1}),
\bfpage{34}
(\byear{2016})
\doiurl{10.3847/0004-637X/830/1/34}
%{\href{https://arxiv.org/abs/1605.03971}{{arXiv:1605.03971}}}
%{[astro-ph.GA]}
\end{barticle}
\endbibitem

%%% 31
\bibitem[\protect\citeauthoryear{Li et~al.}{2016}]{Li2016}
\begin{barticle}
\bauthor{\bsnm{Li}, \binits{T.Y.}},
\bauthor{\bsnm{Wechsler}, \binits{R.H.}},
\bauthor{\bsnm{Devaraj}, \binits{K.}},
\bauthor{\bsnm{Church}, \binits{S.E.}}:
\batitle{Connecting co intensity mapping to molecular gas and star formation in
  the epoch of galaxy assembly}.
\bjtitle{\apj}
\bvolume{817}(\bissue{2}),
\bfpage{169}
(\byear{2016})
\doiurl{10.3847/0004-637X/817/2/169}
%{\href{https://arxiv.org/abs/1503.08833}{{arXiv:1503.08833}}}
%{[astro-ph.CO]}
\end{barticle}
\endbibitem

%%% 32
\bibitem[\protect\citeauthoryear{Bassett et~al.}{2011}]{Bassett2011}
\begin{barticle}
\bauthor{\bsnm{Bassett}, \binits{B.A.}},
\bauthor{\bsnm{Fantaye}, \binits{Y.}},
\bauthor{\bsnm{Hlozek}, \binits{R.}},
\bauthor{\bsnm{Kotze}, \binits{J.}}:
\batitle{Fisher matrix preloaded {\textemdash} fisher4cast}.
\bjtitle{\ijmpd}
\bvolume{20}(\bissue{13}),
\bfpage{2559}--\blpage{2598}
(\byear{2011})
\doiurl{10.1142/S0218271811020548}
%{\href{https://arxiv.org/abs/0906.0993}{{arXiv:0906.0993}}}
%{[astro-ph.CO]}
\end{barticle}
\endbibitem

%%% 33
\bibitem[\protect\citeauthoryear{{DESI Collaboration} et~al.}{2024}]{DESI2024}
\begin{botherref}
\oauthor{\bsnm{{DESI Collaboration}}},
\oauthor{\bsnm{{Adame}}, \binits{A.G.}},
\oauthor{\bsnm{{Aguilar}}, \binits{J.}},
\oauthor{\bsnm{{Ahlen}}, \binits{S.}},
\oauthor{\bsnm{{Alam}}, \binits{S.}},
\oauthor{\bsnm{{Alexander}}, \binits{D.M.}},
\oauthor{\bsnm{{Alvarez}}, \binits{M.}},
\oauthor{\bsnm{{Alves}}, \binits{O.}},
\oauthor{\bsnm{{Anand}}, \binits{A.}},
\oauthor{\bsnm{{Andrade}}, \binits{U. et al.}}:
{DESI 2024 VI: Cosmological Constraints from the Measurements of Baryon
  Acoustic Oscillations}.
arXiv e-prints,
2404--03002
(2024)
\doiurl{10.48550/arXiv.2404.03002}
%{\href{https://arxiv.org/abs/2404.03002}{{arXiv:2404.03002}}}
%{[astro-ph.CO]}
\end{botherref}
\endbibitem

%%% 34
\bibitem[\protect\citeauthoryear{{Dong} et~al.}{2023}]{Dong2023}
\begin{barticle}
\bauthor{\bsnm{{Dong}}, \binits{F.}},
\bauthor{\bsnm{{Park}}, \binits{C.}},
\bauthor{\bsnm{{Hong}}, \binits{S.E.}},
\bauthor{\bsnm{{Kim}}, \binits{J.}},
\bauthor{\bsnm{{Hwang}}, \binits{H.S.}},
\bauthor{\bsnm{{Park}}, \binits{H.}},
\bauthor{\bsnm{{Appleby}}, \binits{S.}}:
\batitle{{Tomographic Alcock-Paczy{\'n}ski Test with Redshift-space Correlation
  Function: Evidence for the Dark Energy Equation-of-state Parameter w $>$ -1}}.
\bjtitle{\apj}
\bvolume{953}(\bissue{1}),
\bfpage{98}
(\byear{2023})
\doiurl{10.3847/1538-4357/acd185}
%{\href{https://arxiv.org/abs/2305.00206}{{arXiv:2305.00206}}}
%{[astro-ph.CO]}
\end{barticle}
\endbibitem

%%% 35
\bibitem[\protect\citeauthoryear{{Planck Collaboration}
  et~al.}{2020}]{PlanckCollaboration2018}
\begin{barticle}
\bauthor{\bsnm{{Planck Collaboration}}},
\bauthor{\bsnm{Aghanim}, \binits{N.}},
\bauthor{\bsnm{Akrami}, \binits{Y.}},
\bauthor{\bsnm{Ashdown}, \binits{M.}},
\bauthor{\bsnm{Aumont}, \binits{J.}},
\bauthor{\bsnm{Baccigalupi}, \binits{C.}},
\bauthor{\bsnm{Ballardini}, \binits{M.}},
\bauthor{\bsnm{Banday}, \binits{A.J.}},
\bauthor{\bsnm{Barreiro}, \binits{R.B.}},
\bauthor{\bsnm{Bartolo}, \binits{N. et al.}}:
\batitle{Planck 2018 results. vi. cosmological parameters}.
\bjtitle{\aap}
\bvolume{641},
\bfpage{6}
(\byear{2020})
\doiurl{10.1051/0004-6361/201833910}
{\href{https://arxiv.org/abs/1807.06209}{{arXiv:1807.06209}}}
{[astro-ph.CO]}
\end{barticle}
\endbibitem

%%% 36
\bibitem[\protect\citeauthoryear{Tojeiro et~al.}{2014}]{Tojeiro2014}
\begin{barticle}
\bauthor{\bsnm{Tojeiro}, \binits{R.}},
\bauthor{\bsnm{Ross}, \binits{A.J.}},
\bauthor{\bsnm{Burden}, \binits{A.}},
\bauthor{\bsnm{Samushia}, \binits{L.}},
\bauthor{\bsnm{Manera}, \binits{M.}},
\bauthor{\bsnm{Percival}, \binits{W.J.}},
\bauthor{\bsnm{Beutler}, \binits{F.}},
\bauthor{\bsnm{Brinkmann}, \binits{J.}},
\bauthor{\bsnm{Brownstein}, \binits{J.R.}},
\bauthor{\bsnm{Cuesta}, \binits{A.J. et al.}}:
% \bauthor{\bsnm{Dawson}, \binits{K.}},
% \bauthor{\bsnm{Eisenstein}, \binits{D.J.}},
% \bauthor{\bsnm{Ho}, \binits{S.}},
% \bauthor{\bsnm{Howlett}, \binits{C.}},
% \bauthor{\bsnm{McBride}, \binits{C.K.}},
% \bauthor{\bsnm{Montesano}, \binits{F.}},
% \bauthor{\bsnm{Olmstead}, \binits{M.D.}},
% \bauthor{\bsnm{Parejko}, \binits{J.K.}},
% \bauthor{\bsnm{Reid}, \binits{B.}},
% \bauthor{\bsnm{Sánchez}, \binits{A.G.}},
% \bauthor{\bsnm{Schlegel}, \binits{D.J.}},
% \bauthor{\bsnm{Schneider}, \binits{D.P.}},
% \bauthor{\bsnm{Tinker}, \binits{J.L.}},
% \bauthor{\bsnm{Magaña}, \binits{M.V.}},
% \bauthor{\bsnm{White}, \binits{M.}}:
\batitle{The clustering of galaxies in the sdss-iii baryon oscillation
  spectroscopic survey: galaxy clustering measurements in the low-redshift
  sample of data release 11}.
\bjtitle{\mnras}
\bvolume{440}(\bissue{3}),
\bfpage{2222}--\blpage{2237}
(\byear{2014})
\doiurl{10.1093/mnras/stu371}
\end{barticle}
\endbibitem

\end{thebibliography}
%\input{paper.bbl}

\end{document}